\documentclass[pdftex,linkcolor=blue,citecolor=blue]{aa} 
\usepackage{array}
\usepackage{diagbox}
\usepackage{makecell}
\usepackage{graphicx}
\usepackage{comment}
\usepackage{orcidlink}

\usepackage{txfonts}
\usepackage{color}

\begin{document} 
    
  \title{\textcolor{black}{Unifying the dynamical classification of early-type galaxies: kinematic deficits in IllustrisTNG versus observations}}
  
    \author{Wenyu Zhong\inst{1}\orcidlink{0009-0005-7009-7914},
          Min Du\inst{1}\orcidlink{0000-0001-9953-0359}\fnmsep\thanks{Min Du and Wenyu Zhong contribute equally to this work. Corr. author: Min Du (E-mail: dumin@xmu.edu.cn)},
          Shengdong Lu\inst{2}\orcidlink{0000-0002-6726-9499},
          Yunpeng Jin\inst{3}\orcidlink{0000-0001-9902-566X}, and
          Kai Zhu\inst{3}\orcidlink{0000-0002-2583-2669}         
          }

   \institute{Department of Astronomy, Xiamen University, Xiamen, Fujian 361005, China
   \and
    \textcolor{black}{Institute for Computational Cosmology, Department of Physics, University of Durham, South Road, Durham, DH1 3LE, UK} 
     \and 
     Department of Astronomy, Westlake University, Hangzhou, Zhejiang 310030, China
             }
    
   \date{Received 2025/09/19; accepted 2026/01/30}
 
  \abstract
    {}
    {This study aims to make a comparative analysis of galaxy kinematics using IllustrisTNG simulations and integral-field spectroscopy (IFS) observations.}
    {We identify 2,342 early-type galaxies (ETGs) from the TNG100 simulation and 236 ETGs from the TNG50 simulation to compare with those from MaNGA and ATLAS$^{\rm 3D}$ surveys. For these systems, we measure key kinematic parameters widely employed in both observational and simulation-based studies, including the intrinsic spin parameter \(\lambda_{R,\rm intr}\) (it indicates that the $\lambda_{R}$ parameter is measured when the galaxy is viewed edge-on), the cylindrical rotational energy fraction \(\kappa_{\rm rot}\), and structural mass ratios such as the spheroid mass fraction \(f_{\rm spheroid}\) and the stellar halo mass fraction $f_{\rm halo}$.}
    {This study performs a comparative kinematic analysis of early-type galaxies using IllustrisTNG simulations and IFS data from MaNGA and ATLAS$^{\rm 3D}$. We demonstrate that standard classifiers—the $\lambda_{R}(R_e) = 0.31\sqrt{\varepsilon}$ relation and  $\overline{k_5}$ coefficient (the higher-order term of the Fourier decomposition of velocity fields) —fail to align with kinematic bimodality. Revised thresholds are proposed: the spin $\lambda_{R,\rm{intr}}$ $\sim 0.4$, the ratio of rotation energy $\kappa_{\rm rot} \sim 0.5$, and the mass fraction of a spheroid component $f_{\rm spheroid} \sim 0.6$. It provides a universal threshold that classifies all kinds of galaxies into rotation-dominated (fast rotators) and random motion-dominated (slow rotators) cases. Scaling relations derived from TNG enable estimation of $\kappa_{\rm rot}$ and $f_{\rm spheroid}$ from observations. TNG simulations exhibit a bimodality deficit, characterized by a lack of fast rotators and suppressed $\lambda_{R,\rm{intr}}$, attributable to excess galaxies with intermediate rotation and high spheroid/stellar halo mass. A novel method for estimating stellar halo mass fractions from IFS kinematics is introduced, though significant uncertainties persist.}
   {}


  \titlerunning{Classification of ETGs}
   \maketitle

\section{Introduction}

    The consensus holds that early-type galaxies (ETGs) are \textcolor{black}{generally} quiescent systems that underwent early cessation of star formation, resulting in their current red optical colors and minimal reserves of cold gas and dust \citep{1999Galaxy}. Morphologically, ETGs appear symmetric and predominantly elliptical, with shapes ranging from near-circular to highly elongated \citep{2005ARA&A..43..581S}. Classified within Hubble’s framework \citep{1926ApJ....64..321H,1936rene.book.....H}, ETGs comprise two main types, \textcolor{black}{i.e.}, elliptical and lenticular \textcolor{black}{(S0)} galaxies. Elliptical galaxies are smooth triaxial ellipsoids devoid of spiral arms or nebulae. Their surface brightness follows de Vaucouleurs' law \citep{1948AnAp...11..247D}, featuring concentrated central peaks and sharp outward declines \citep{1977ApJ...218..333K}. Lenticular galaxies retain disk-like structures but lack spiral arms, often exhibiting central bars or rings \citep[e.g.,][]{1970ApJ...160..831S,1976ApJ...206..883V,2003AJ....125.2936G}. \par

    Integral Field Spectroscopy (IFS) technology enables the classification of dust-unaffected ETGs into ``Fast Rotators (FRs)'' and ``Slow Rotators (SRs)'' \textcolor{black}{based on projected kinematic properties} \citep{2004MNRAS.352..721E,2007MNRAS.379..401E,2011MNRAS.414..888E}. This division originated from qualitative observations in the SAURON survey of 48 ETGs \citep{2004MNRAS.352..721E}, which revealed a dichotomy: galaxies exhibited either regular rotation patterns or complex/irregular stellar velocity maps, \textcolor{black}{and are defined as regular rotators (RR)/non-regular rotators (NRR).} Quantification of kinematic regularity is achieved through harmonic term analysis (e.g., \textcolor{black}{the higher-order term of the Fourier decomposition of velocity fields $\overline{k_5}$} \textcolor{black}{\citep{2006MNRAS.366..787K, 2008MNRAS.390...93K}, see details in Section \ref{KinematicParameter}}), distinguishing regular rotators from non-regular rotators \citep{2011MNRAS.414.2923K, 2016ARA&A..54..597C}. The correspondence between RR/FR and NRR/SR led to the adoption of the spin parameter \textcolor{black}{\textcolor{black}{$\lambda_{R}$} \citep{2011MNRAS.414..888E, 2007MNRAS.379..401E}}, with SRs defined by \textcolor{black}{$\lambda_{R} < 0.31\sqrt{\varepsilon}$} within the effective radius $R_e$, \textcolor{black}{where $\varepsilon$ is the ellipticity}. FRs display high \textcolor{black}{\textcolor{black}{$\lambda_{R}$}} and $\varepsilon$, resembling disk galaxies, while SRs feature weak rotation \textcolor{black}{($\lambda_{\rm R} \lesssim 0.15$)}. IFS surveys like SAURON \citep{2002MNRAS.329..513D}, $\text{ATLAS}^{\rm 3D}$ \citep{2011MNRAS.413..813C}, CALIFA \citep{2012A&A...538A...8S}, SAMI \citep{2012MNRAS.421..872C,2015MNRAS.447.2857B}, MASSIVE \citep{2014ApJ...795..158M}, and MaNGA \citep{2015ApJ...798....7B} provide spatially resolved kinematic maps, facilitating studies of ordered versus disordered motion via \textcolor{black}{\textcolor{black}{$\lambda_{R}$}-$\varepsilon$} diagnostics. However, \textcolor{black}{the robustness of this classification has not been} systematically examined in numerical simulations. And the correlation between \textcolor{black}{$\overline{k_5}$} and other kinematic parameters, e.g., the relative importance of cylindrical rotation energy \citep{2010MNRAS.409.1541S} that have been widely used in simulations, is poorly quantified.\par

    Recent cosmological hydrodynamical simulations, e.g., EAGLE \citep{2015MNRAS.446..521S,2015MNRAS.450.1937C}, IllustrisTNG \textcolor{black}{\citep[TNG hereafter, e.g.,][]{2018MNRAS.475..624N}}, \textcolor{black}{Horizon-AGN \citep{2014MNRAS.444.1453D,2016MNRAS.463.3948D}, New-Horizon \citep{Dubois_2021}, \textcolor{black}{Magneticum \citep{2025arXiv250401061D},} and SIMBA \citep{2019MNRAS.486.2827D}}, have made remarkable progress that successfully reproduces realistic visual morphologies and properties in many aspects. \textcolor{black}{These remarkable progresses allow} statistical studies of galaxies in a fully self-consistent cosmological environment. On one hand, a systematic comparison between IFS results and kinematic properties in simulations is crucial for constraining the potential shortage of simulations. On the other hand, numerical simulations can be used to examine the kinematic models and classification of SRs and FRs. \textcolor{black}{It must be pointed out that, in addition to being able to reproduce fairly well the morphological properties observed, modern cosmological hydrodynamic simulations – including the TNG suite – demonstrate a certain degree of reliability in reproducing the observed kinematic properties. \citet{2024MNRAS.528.2326S} first established that the EAGLE simulation successfully models the stellar mass fraction variation across orbital families as functions of stellar mass and spin parameter at \(z = 0\), consistent with SAMI observations. This alignment extends to the correlation between specific angular momentum $j_*$ and stellar mass $M_*$, where \citet{2017MNRAS.464.3850L} reported reasonable agreement between EAGLE results and \({\rm ATLAS}^{\rm 3D}\) data. Similarly, analyses of the {Magneticum Pathfinder} simulations reveal highly congruent \(\lambda_R(R_e)\) and \(\varepsilon\) distributions for ETGs compared to \({\rm ATLAS}^{\rm 3D}\), CALIFA, SAMI, and SLUGGS surveys \citep{2018MNRAS.480.4636S}. Further validation by \citet{2020MNRAS.493.3778S} confirms that Magneticum's stellar spin parameter profiles closely match observations from SLUGGS \citep{2014ApJ...791...80A,2017MNRAS.464.4611F} and ePN.S \citep{2018A&A...618A..94P}.} Building on CALIFA's kinematic "Hubble sequence" derived from stellar orbit-circularity distributions, \citet{Xu_2019} analyzed $z=0$ TNG galaxies to explore connections between stellar orbital compositions and fundamental galaxy properties. They found TNG100 broadly reproduces observed orbital-component fractions and their stellar-mass dependencies. \textcolor{black}{Similarly,} \citet{2025A&A...699A.320Z} further presented that TNG galaxies match well with the luminosity fractions of cold, warm, and hot orbital components \textcolor{black}{of CALIFA galaxies that are extracted via performing the orbit-superposition Schwarzschild modeling \citep[e.g.,][]{1979ApJ...232..236S,2004ApJ...602...66V,2008MNRAS.385..647V,2018NatAs...2..233Z,2018MNRAS.479..945Z,2018MNRAS.473.3000Z}}. \textcolor{black}{Moreover,
    IllustrisTNG have reproduced galaxies with unusual
    structural and kinematic properties, such as low surface brightness galaxies \citep{2018MNRAS.480L..18Z}, and jellyfish galaxies \citep{2019MNRAS.483.1042Y}.} Therefore, \textcolor{black}{to a certain extent}, TNG simulations can be employed for further research on the kinematic parameters and properties of galaxies. \par 

    Structural mass ratios derived from kinematic decomposition and cylindrical rotational properties are able to characterize galaxy dynamical properties and their underlying formation histories. However, observational measurement of spins, $\kappa_{\rm rot}$, and precise structural mass ratios are challenging. For simulations, kinematic phase-space analysis enables robust decomposition: the \texttt{auto-GMM} method, an unsupervised algorithm developed by \cite{2019ApJ...884..129D}, identifies structures such as cold disks, warm disks, bulges, and halos in TNG galaxies by clustering stellar particles in a 3D kinematic phase space of normalized azimuthal angular momentum (circularity), non-azimuthal angular momentum, and binding energy. Galaxies with minimal stellar halos exhibit tight scaling relations among $j_*$, $M_*$, size, and metallicity owing to their quiescent merger histories \citep{2022ApJ...937L..18D, 2024A&A...686A.168D, 2024A&A...689A.293M}. Despite the \texttt{auto-GMM}'s success in isolating intrinsic structures formed by distinct mechanisms—such as separating stellar halos from bulges \citep{2020ApJ...895..139D, 2021ApJ...919..135D}—the kinematic decomposition cannot be directly applied to observational datasets. Consequently, linking kinematically derived structures as well as galactic spin to IFS observables is critical. Moreover, while $\kappa_{\rm rot}=0.5$ or a spheroid mass ratio of 0.5 serves as a common threshold for classifying elliptical versus disk galaxies \textcolor{black}{\citep[e.g.,][]{Zhao_2020,2023ApJ...949L..14F,2024MNRAS.533...93P,2024MNRAS.527.3366K}}, its relationship to kinematically distinct FRs and SRs in IFS surveys remains unclear.\par

    In this paper, we \textcolor{black}{perform} a comparative analysis of galaxy kinematics using TNG simulations and IFS data from the MaNGA and ATLAS\({}^{\text{3D}}\) surveys, with a focus on classifying early-type galaxies into FRs and SRs, \textcolor{black}{as well as interpreting the definitions of FRs and SRs}. The structure of this paper is organized as follows. Section \ref{observation ETGs} describes the ETGs sample selection and data extraction from IFS observations. Section \ref{TNG kinematic} describes mock observations and sample selection of TNG simulation galaxies. Section \ref{KinematicParameter} demonstrates the kinematic parameters quantifying rotation properties of galaxies. Section \ref{label_scaling_krot_spheroid} investigates the relation between different kinematic parameters. Section \ref{Kinematic classification} examines the bimodality and the classification of slow and fast rotators in kinematics. Section \ref{section rotation} compares the radial distribution of kinematic properties between TNG simulations and observations. \textcolor{black}{Our conclusions are discussed and summarized in Sections \ref{evalute_cosmological_simulation} and \ref{conclusion}, respectively.}\par

\section{Sample selection and data extraction from IFS observations}\label{observation ETGs}

\subsection{Data extraction of color, morphology, and rotation}

    This study compares the kinematic properties of simulated TNG galaxies with IFS measurements from the ATLAS$^{\text{3D}}$ and MaNGA surveys for ETGs. $\text{ATLAS}^{\text{3D}}$\footnote{http://www-astro.physics.ox.ac.uk/atlas3d/} is a complete, multi-wavelength, volume-limited survey of nearby ETGs, \textcolor{black}{and its final sample encompasses 260 nearby ETGs.} The Sloan Digital Sky Survey-IV \citep[SDSS-IV:][]{2000AJ....120.1579Y, 2006AJ....131.2332G, 2017AJ....154...28B} Mapping Nearby Galaxies at Apache Point Observatory (MaNGA\footnote{https://www.sdss4.org/surveys/manga}, \citealt{2015ApJ...798....7B, 2016AJ....152..197Y}) is an IFS survey designed to obtain spectral measurements across the surfaces of approximately 10,000 nearby galaxies. The data of color, morphology, and stellar mass of our sample are extracted as follows.\par 
    
    {Color $g-r$} $-$ The $g-r$ colors for MaNGA galaxies and most ATLAS$^{\text{3D}}$ galaxies are obtained from the NASA-Sloan Atlas \textcolor{black}{\footnote{http://www.nsatlas.org} (NSA) catalog \citep{2007AJ....133..734B,2011AJ....142...31B}}. For the subset of ATLAS$^{\text{3D}}$ galaxies not included in the NSA catalog, we use $B-V$ colors corrected for Galactic extinction from the Hyperleda\footnote{http://leda.univ-lyon1.fr/} catalog \citep{2014A&A...570A..13M}. These $B-V$ colors are then converted to $g-r$ colors using the relation provided in \cite{2020A&A...641A..60P}. \par

    {Size $R_e$} $ - $ The effective radius $R_e$ of the ATLAS$^{\text{3D}}$ sample is adopted from the Table 1 in \cite{2013MNRAS.432.1709C}. The data of MaNGA is from \cite{2023MNRAS.522.6326Z}. MaNGA provides spatially resolved spectra, covering radial ranges out to $1.5R_e$ for the Primary+ sample (about two-thirds of the total sample) and out to $2.5R_e$ for the Secondary sample (about one-third of the total sample). \par

    {Stellar mass $M_{*}$} $-$ \textcolor{black}{The stellar masses for the $\text{ATLAS}^{\rm 3D}$ and MaNGA sample are derived using a color–mass relation assuming the \cite{2003PASP..115..763C} initial mass function.} For the ATLAS\(^{\text{3D}}\) dataset, total absolute \(K\)-band luminosities (\(L_K\)) are adopted directly from \citet{2011MNRAS.413..813C}. For the MaNGA dataset, \(L_K\) values originate from \cite{2018MNRAS.477.4711G}. Both datasets use luminosities sourced from the 2MASS extended source catalog \citep{2003AJ....125..525J}, with corrections applied for Galactic extinction. These luminosities are further calibrated to address sky-background over-subtraction artifacts using \(L_{K_{\text{corr}}} = 1.07L_K + 1.53\) as established by \cite{2013ApJ...768...76S}. Finally, stellar masses are calculated via
    \(
    \log_{10} (M_{*}/M_\odot) = 10.39 - 0.46(L_{K_{\text{corr}}} + 23)
    \)
    \citep{2019MNRAS.484..869V}, where \(M_*\) is consistent with that derived by the Jeans Anisotropic Models (JAM) dynamical methodology \citep{2013MNRAS.432.1862C,2013ApJ...778L...2C}.\par 

    \begin{figure}[t]
        \centering	\includegraphics[width=0.99\columnwidth]{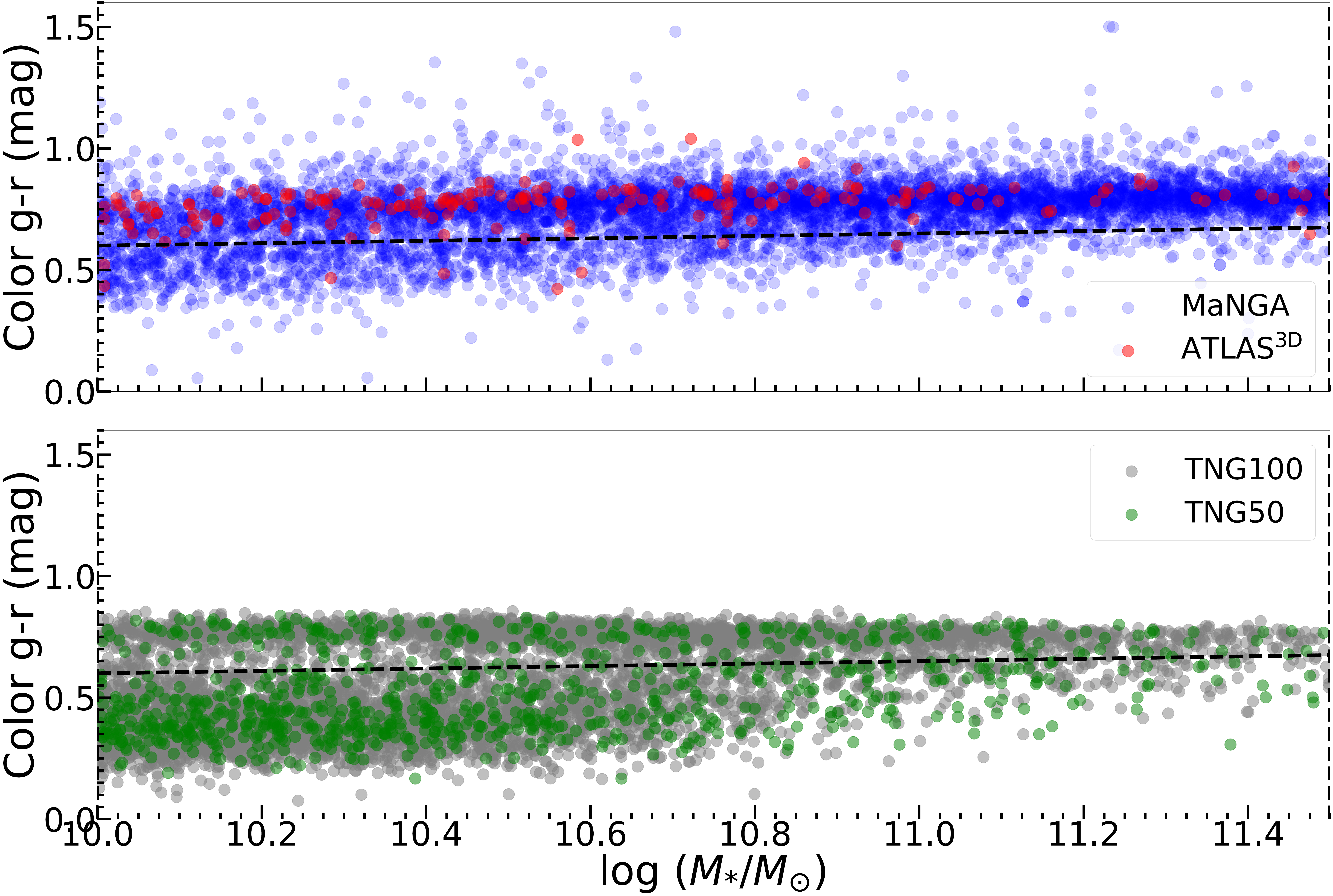}
            \caption{The $g-r$ color-stellar mass diagram. \textcolor{black}{The upper and lower panels of the figure display observational data (from MaNGA and  $\text{ATLAS}^{\rm 3D}$) and simulated data (from TNG50 and TNG100), respectively, encompassing all samples within the mass range of $10^{10}-10^{11.5}M_{\odot}$}. Red sequence ETGs are selected above the dashed line.} 
        \label{fig:obs_TNG_grindex}
    \end{figure}
       
    {Ellipticity $\varepsilon$} $-$ For ATLAS$^{\text{3D}}$ galaxies, ellipticities within 1$R_e$ are  from \cite{2011MNRAS.414..888E} Table B1. \textcolor{black}{Among} the 260 galaxies analyzed, 17 with prominent bars have ellipticity measurements specifically at 2.5–3$R_e$ to avoid biased inner regions \textcolor{black}{\citep{2011MNRAS.414..888E,2020A&A...641A..60P}}. 
    \textcolor{black}{The ellipticities of MaNGA galaxies are obtained from \cite{2023MNRAS.522.6326Z}, which are uniformly measured at $1R_e$.}\par 
       
    {Spin spectroscopic parameter \textcolor{black}{\textcolor{black}{$\lambda_{R}$}}} $-$ The spin spectroscopic parameter \textcolor{black}{\textcolor{black}{$\lambda_{R}$}} is calculated across different surveys using slightly varying integration areas. In the ATLAS$^{\text{3D}}$ survey, \citet{2011MNRAS.414..888E} use circular apertures with a radius of $1R_e$, whereas the MaNGA survey integrates over a half-light ellipse \citep{2023MNRAS.522.6326Z}. This ellipse covers an area equivalent to a circle of radius $1R_e$, with a semi-major axis of \textcolor{black}{$R_e /\sqrt{1-\varepsilon}$}. \textcolor{black}{Meanwhile, we adopt elliptical apertures for all $\lambda_R$ calculations in this paper.}\par

     $V_{\mathrm{rot}}/\sigma$ {profiles} $-$ The $V_{\mathrm{rot}}/\sigma$ profiles for both the $\text{ATLAS}^{\text{3D}}$ and MaNGA galaxies are \textcolor{black}{calculated} as the ratio of the rotation velocity $V_{\rm rot}(R)$ \textcolor{black}{(see Equation (\ref{kinemetry_eq})}) and the azimuthally averaged velocity dispersion $\sigma(R)$ at a given in \textcolor{black}{elliptical radial bins}  \citep{2020A&A...641A..60P}. For the $\text{ATLAS}^{\text{3D}}$ sample, kinematic maps are adopted from \citet{2004MNRAS.352..721E} and \citet{2011MNRAS.413..813C}. For the MaNGA galaxies, publicly available stellar kinematics from the SDSS DR17 \footnote{https://www.sdss4.org/dr17/manga/manga-data/} \citep{2022ApJS..259...35A} are used. We then apply KINEMETRY \footnote{https://www.aip.de/en/members/davor-krajnovic/kinemetry/} to fit the velocity and velocity dispersion fields separately, deriving radial profiles of $V_{\mathrm{rot}}$ and $\sigma$ for these galaxies.\par 
     
    {Inclination angle} $i$ $-$  Inclination angles of ATLAS$^{\text{3D}}$ and MaNGA galaxies are adopted from \cite{2013MNRAS.432.1709C} and \citet{2023MNRAS.522.6326Z}, respectively. 

\subsection{Sample selection of early-type galaxies}\label{sec:obs_sample}

    The ATLAS$^{\text{3D}}$ survey selected a sample of bright ETGs with high completeness within a distance of $<42$ Mpc ($z\lesssim 0.01$) and a sky declination $\delta$ satisfying $|\delta - 29^{\circ}| < 35^{\circ}$. Additionally, these galaxies were required to have an absolute $K$-band magnitude brighter than $-21.5$ mag ($M_{*} \gtrsim 6\times 10^9 M_{\odot}$), and away from the dusty region near the Galaxy equatorial plane with latitude $|b| < 15^{\circ}$. ETGs were identified morphologically by excluding galaxies with visible spiral structures, resulting in a sample similar to galaxies on the red sequence \citep{2011MNRAS.413..813C}. The ATLAS$^{\text{3D}}$ ETG sample comprises 68 elliptical galaxies and 192 lenticular galaxies. The final sample comprises 260 nearby ETGs, with their kinematic data acquired using the SAURON IFS instrument \citep{2001MNRAS.326...23B} mounted on the William Herschel Telescope. The SAURON IFS has a field of view of approximately $33^{''} \times 41^{''}$ and is equipped with a square microlens array of $0.94^{''} \times 0.94^{''}$. For the $\text{ATLAS}^{\text{3D}}$ galaxies, \cite{2011MNRAS.413..813C} employed a uniform analysis pipeline to analyze the spectral datacubes, adopting a minimum signal-to-noise (S/N) threshold of 40 for adaptive binning \citep{2003MNRAS.342..345C}. Stellar kinematic was extracted using the ppxf\footnote{\textcolor{black}{https://pypi.org/project/ppxf/}} software \citep{2004PASP..116..138C} in conjunction with the MILES\footnote{https://research.iac.es/proyecto/miles/} stellar template library \citep{2006MNRAS.371..703S}\par

    \begin{figure}[t]
       \centering        \includegraphics[width=0.99\columnwidth]{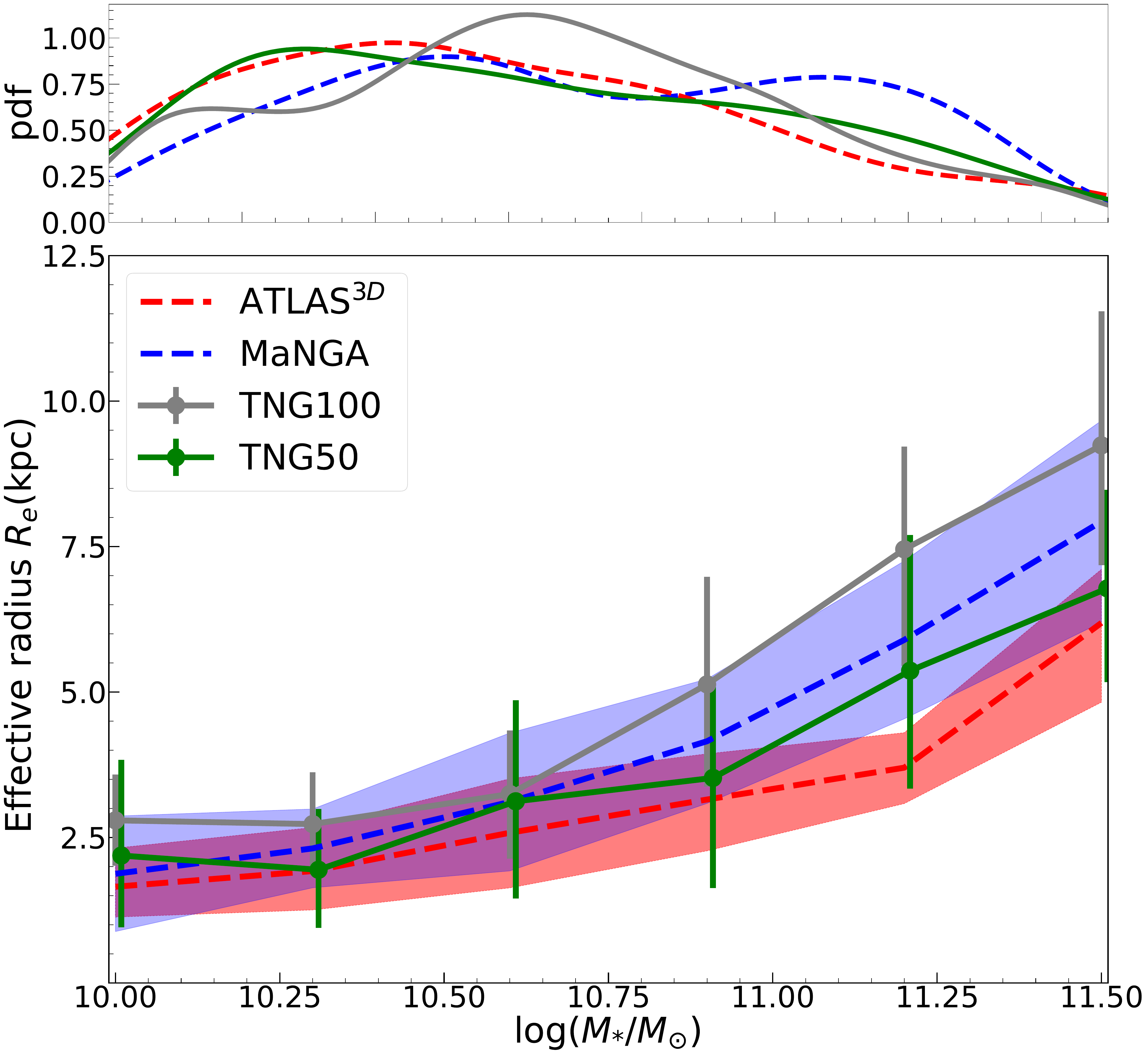}
    	\caption{Mass-size relation. \textit{Top}: stellar mass distribution function of the chosen sample of ETGs from observational data and the TNG simulations. \textit{Bottom}: Circularized effective radii $R_e$, of the selected ETG samples in the TNG100 and TNG50 simulations as a function of stellar mass, and comparison with observations (ATLAS$^{\mathrm{3D}}$ and MaNGA). Both TNG50 \textcolor{black}{(green solid line)} and TNG100 \textcolor{black}{(gray solid line)} galaxies share comparable stellar masses and sizes with the ATLAS$^{\mathrm{3D}}$ \textcolor{black}{(red dashed line)} and MaNGA \textcolor{black}{(blue dashed line)} galaxies.  \textcolor{black}{The lines and the points represent the median values, with colored regions and error bars indicating the range from the 16th to 84th percentile of each distribution. \textcolor{black}{To avoid overlap, we shift the error bars slightly for clarity.}}}
    	\label{fig:mass_size_relation}
     \end{figure} 
    The MaNGA survey targets galaxies within a redshift range of \(0.01 < z < 0.15\) and a stellar mass range of \(10^9\,M_\odot\) to \(10^{12}\,M_\odot\), achieving a spatial resolution of \(2.5^{\prime\prime}\). For the MaNGA sample, stellar kinematics are derived from spectral data cubes using the Data Analysis Pipeline (DAP; \citet{2019AJ....158..160B, 2019AJ....158..231W}) with a subset of the MILES stellar library \citep{2006MNRAS.371..703S, 2011A&A...532A..95F}. Prior to kinematic extraction, the spectra are Voronoi-binned \citep{2003MNRAS.342..345C} to a target S/N of 10 to ensure reliable stellar velocity dispersion measurements. The final SDSS DR17 includes over 10,000 galaxy data cubes after the removal of non-galaxy targets. We \textcolor{black}{only select} datacubes with good quality\footnote{https://manga-dynpop.github.io/pages/data\underline{~}access} (drp3qual=$1$) as determined by MaNGA's Data Reduction Pipeline \citep[DRP;][]{2016AJ....152...83L,2016AJ....152..197Y}, and impose additional criteria: redshift $z<0.05$ (\textcolor{black}{The sample is considered to be approximately complete.}) and stellar masses between $10^{10} M_\sun$ and $10^{11.5} M_\sun$. This yields a final sample of 3,100 MaNGA galaxies that is comparable with the TNG sample.\par 

     \textcolor{black}{We select ETGs using \textcolor{black}{the criterion} $(g-r) \ge 0.05 \log_{10}(M_*/M_\odot) + 0.1$ mag \citep{2020A&A...641A..60P} and limited the sample stellar mass range to $10^{10} \leq M_{*} \leq 10^{11.5} M_{\odot}$}, yielding 193 ATLAS$^{\mathrm{3D}}$ (red points) ETGs, shown in the upper panel of Figure \ref{fig:obs_TNG_grindex}. \textcolor{black}{Relying solely on color for classification proves inadequate for MaNGA galaxies (blue points), as dust reddening in spiral galaxies can significantly alter their observed colors. To address this limitation, we supplement $g-r$ color cuts with morphological classifications from the MaNGA Deep Learning catalog\footnote{https://www.sdss4.org/dr15/data\_access/value-added-catalogs/?vac\_id=manga-morphology-deep-learning-dr15-catalog},} which utilizes a convolutional neural network trained on data from \citet{2010ApJS..186..427N}. \textcolor{black}{Therefore, to select ETGs from the MaNGA data, two criteria must be met simultaneously: first, these galaxies should have relatively red colors; second, they should be classified as ETGs according to the Deep Learning catalog.} Adopting the methodology of \citet{2022MNRAS.509.4024D}, we select early-type galaxies using the same criteria. This approach identifies 599 elliptical and 334 lenticular galaxies.\par 

\section{\textcolor{black}{Sample selection and mock observations} of IllustrisTNG simulation galaxies}\label{TNG kinematic}

    The TNG simulations use  $\Lambda$CDM setup \textcolor{black}{with the cosmological parameters from} the Planck survey \citep{2016A&A...594A...1P}, and it exploits all the advantages of the unstructured moving-mesh hydrodynamic method {\tt Arepo}\footnote{https://gitlab.mpcdf.mpg.de/vrs/arepo} \citep{2010MNRAS.401..791S}, but improves the numerical methods, the subgrid physical
    model, and the recipe for galaxy feedback both from stars and AGN. In particular, TNG is equipped with a novel dual mode (thermal and kinetic) AGN feedback that shapes and regulates the stellar component within massive systems, maintaining a realistic gas fraction \citep{2017MNRAS.465.3291W}. Also, the feedback model from galactic winds has been improved to have better representation of low- and
    intermediate-mass galaxies \citep{2018MNRAS.473.4077P}.\par

    The TNG suite comprises three runs using different simulation volumes and resolutions, namely TNG50, TNG100, and TNG300 \textcolor{black}{\citep{2018MNRAS.475..624N,2018MNRAS.477.1206N,2018MNRAS.480.5113M,2018MNRAS.475..648P,2019MNRAS.490.3196P,2018MNRAS.475..676S}}. Each run starts at $z = 127$ using the Zeldovich approximation and evolves down to $z = 0$. In this study, we use TNG50-1 (hereafter, simply TNG50) and TNG100-1 (hereafter, simply TNG100) runs. TNG50 provides \textcolor{black}{a large number of galaxies} for statistical analyses and a \textcolor{black}{``zoom-in''}-like resolution. TNG50 includes $2 \times 2160^3$ initial resolution elements in a $\sim$50  comoving Mpc box, \textcolor{black}{with the} baryon mass resolution of 8.5 $\times$ $10^{4} M_{\odot}$ \textcolor{black}{and} a gravitational softening radius $r_{\mathrm{soft}}$ for stars of about 0.3 kpc at $z = 0$. Dark matter is resolved with particles of mass 4.5 $\times$ $10^{5}M_{\odot}$. Meanwhile, the minimum gas softening reaches 74 comoving parsecs. \textcolor{black}{TNG100 includes approximately $2 \times 1820^3$ resolution elements in a \textcolor{black}{$\sim$110 comoving Mpc} box. The dark matter (DM) and baryonic mass resolutions of TNG100 are $m_{\rm DM} = 7.5 \times 10^6 M_{\odot}$ and $m_b = 1.4 \times 10^6 M_{\odot}$, respectively \citep{2003MNRAS.339..289S}.} The softening length employed for TNG100 for both the DM and stellar components is $\epsilon=0.74$ kpc, while an adaptive gas gravitational softening is used, with a minimum $\epsilon_{\rm gas, min}= 0.185$ kpc. TNG50 thus has roughly 15 times better mass resolution, and 2.5 times better spatial resolution, than TNG100 \citep{2019MNRAS.490.3234N}\par
    
    \textcolor{black}{The identification of galaxies in the simulations is performed} using the Friends-of-Friends (FoF) group finding algorithm \citep{1985ApJ...292..371D} and the {\tt SUBFIND} algorithm \textcolor{black}{\citep{2001ApJ...549..681S,2009MNRAS.399..497D}}. \textcolor{black}{The central galaxy (subhalo) is the first (most massive) subhalo of each FoF group, while the other galaxies within the FoF group are its satellites.} All components-including gas, stars, dark matter, and black holes-that are gravitationally bound to a single galaxy \textcolor{black}{are associated with their host subhalo.} Each galaxy is positioned at the minimum of its respective gravitational potential well. This study utilizes galaxies from snapshot $\#99$ (corresponding to $z=0$) in both TNG50 and TNG100. \par

\subsection{Sample selection of early-type galaxies}\label{sample select sec}

    \citet{2018MNRAS.475..624N} showed that TNG simulations reproduce realistic galaxy colors matching SDSS observations \citep{2001AJ....122.1861S}. Redder galaxies in TNG exhibit typical \textcolor{black}{ETG-like} properties: suppressed star formation, low gas fractions, \textcolor{black}{higher metallicities}, and older stellar populations. \textcolor{black}{Firstly, we select all galaxies with stellar masses within the range of $10^{10.0} \leq M_{*} \leq 10^{11.5} M_{\odot}$, including both central and satellite galaxies.} Then we identify ETGs using the color-mass diagram (Figure~\ref{fig:obs_TNG_grindex}, lower panel), \textcolor{black}{applying the same color selection criteria (dashed lines) and  the same stellar mass range as we use for observations.} While TNG color measurements ignore dust effects, we validated our ETG selection using a deep learning classifier \citep{HuertasCompany2019} trained on \citet{2010ApJS..186..427N} data\footnote{https://www.tng-project.org/data/docs/specifications/\#sec5r}. This morphological analysis confirmed that 85\% of color-selected TNG100 galaxies are genuine ETGs (1,693 E/S0 systems), with the remaining 15\% being spiral galaxies. Tests show this contamination negligibly impacts our results, justifying our color-only selection approach for TNG samples.\par 

    In addition, \textcolor{black}{we exclude galaxies where the effective radius $R_e$ fell below twice the gravitational softening length ($R_e < 2 r_\mathrm{soft}$) where $r_{\mathrm{soft}}$ is 0.74 kpc and 0.288 kpc in TNG100 and TNG50, respectively. This criterion will eliminate 5 objects (2.0\% of the TNG50 parent sample; N=244) and 245 objects (9.4\% of the TNG100 parent sample; N=2602).} \textcolor{black}{We further exclude merger-contaminated galaxies through visual inspection of stellar density maps.} After applying these criteria, our final ETG sample consists of 2342 and 236 galaxies from TNG100 and TNG50, respectively. \par

     \begin{figure}[t]
       \centering        \includegraphics[width=0.99\columnwidth]{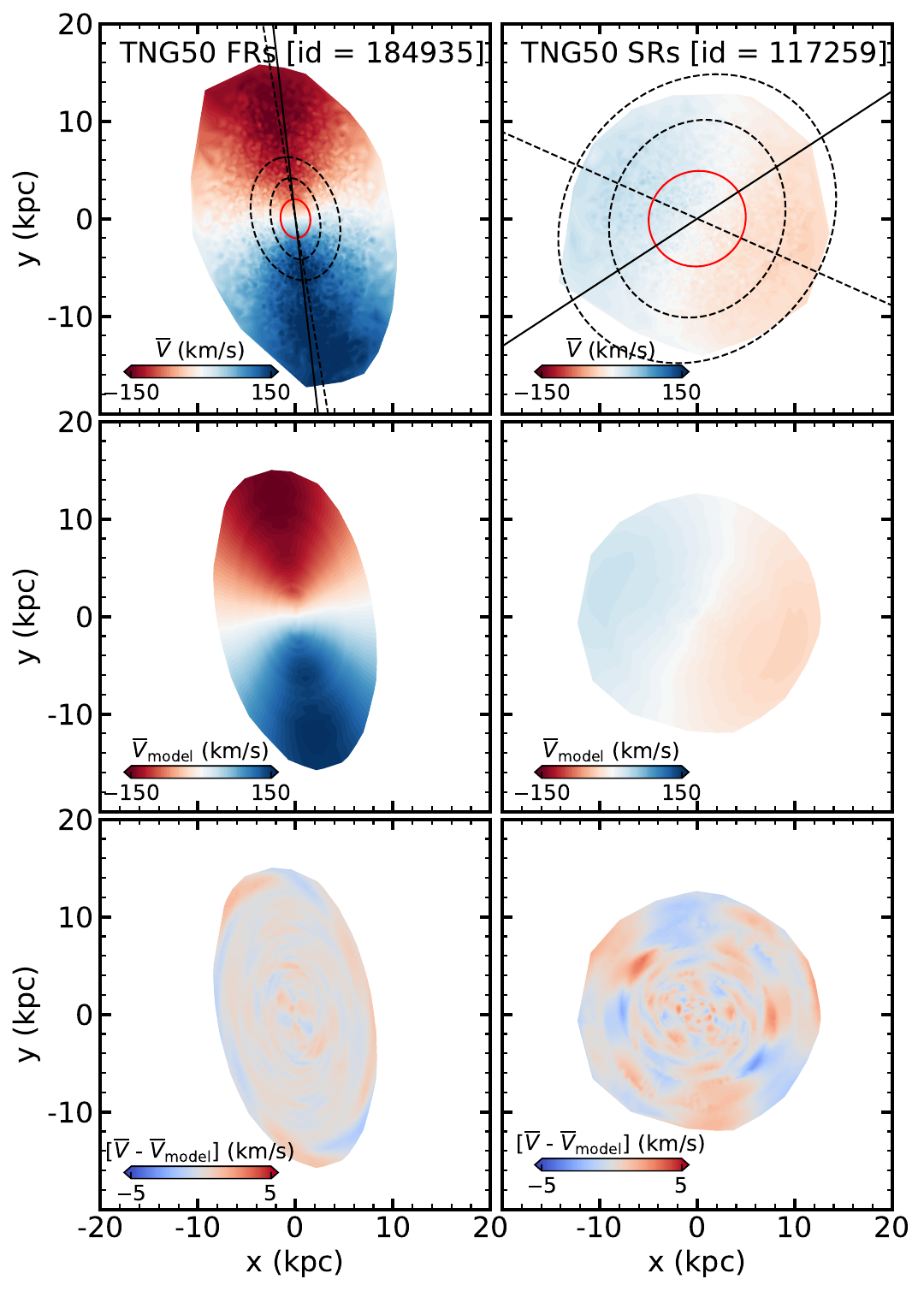}
       \caption{Velocity maps and models of examples of a fast rotator (left) and a slow rotator (right). It shows raw data at the top, a reconstructed velocity field employing a pure cosine law along ellipses in the middle, and residuals depicting deviations at the bottom. Our visualization emphasizes structural features through isodensity ellipses spaced at regular $R_e$ intervals from $R_e$ to $3R_e$. The major axis orientations - with solid lines representing photometrically-derived morphological axes and dashed lines indicating kinematically-determined axes - are both measured at $R_e$.}
    	\label{fig:FRs_SRs_kinemetry_model}
    \end{figure} 

    The $\text{ATLAS}^{\text{3D}}$ parent sample, which includes 260 ETGs and 611 spiral/irregular galaxies at \(z \lesssim 0.01\), has been confirmed as a complete and representative sample of the nearby galaxy population, with an ETG fraction of approximately 30\% (260/871) \citep{2011MNRAS.413..813C}. In the TNG simulations, our selection criteria yield ETG fractions of $27\%$ (236/873) in TNG50 and $35\%$ (2342/6369) in TNG100. By comparison, \textcolor{black}{an observational sample (MaNGA)} at \(z<0.05\) (Section \ref{sec:obs_sample}) classifies $23\%$ of galaxies as ETGs, a value closely aligned with TNG50. A separate local sample (\(0.05 < z < 0.1\), \(g < 16\) mag) reports a higher ETG fraction of 33\% (14,034 galaxies) \textcolor{black}{\citep{2010ApJS..186..427N}}. TNG100 appears to reproduce a more consistent overall fraction of ETGs, while the discrepancy between TNG100 and TNG50 suggests that TNG50 may underestimate quenching efficiency. \textcolor{black}{Meanwhile, in terms of spatial coverage and environmental representation, TNG50 broadly aligns with the ${\rm ATLAS}^{\rm 3D}$ survey in sampling low-density cosmological regimes \citep[e.g.,][]{2011MNRAS.416.1680C}, while TNG100 corresponds to MaNGA's cosmological volume ($z<0.05$) which incorporates higher-density environments \citep[e.g.,][]{2017MNRAS.465..688G,2017ApJ...851L..33G}. Shown in Figure~\ref{fig:mass_size_relation}, the stellar mass distributions of TNG50 and TNG100 galaxies show no clear differences attributable to their cosmological environments. And these environmental differences between simulation pairs do not fully account for observational-simulation discrepancies we found (see details in Section \ref{sec_bimodality}).}\par

\subsection{Mock images and data extraction}
    \subsubsection{Mass and size} 
    We define the total stellar mass $M_{*}$
    as the total bound stellar mass of the galaxy. \textcolor{black}{That is, the calculation of the total stellar mass utilizes the complete set of particle data associated with galaxies and subhalos, encompassing all gravitationally bound particles identified through the {\tt SUBFIND} algorithm.}
    The \textcolor{black}{effective radius} $R_e$ is estimated by the circularized effective radius in a similar way to that of observations, where the area of the ellipse that encloses half of $M_{*}$ is equated to \( \pi R_e^{2} \). \par 

    The Multi-Gaussian Expansion (MGE) fitting is conducted using the Python package MGEFIT\footnote{Version 5.0.13, available at https://pypi.org/project/mgefit/} \citep{2002MNRAS.333..400C} on the 2-dimentional stellar mass map \textcolor{black}{within an area of 60 kpc $\times$ 60 kpc.} The stellar surface density in the MGE fitting is expressed as:
    \begin{equation}
	\Sigma(x^{'}, y^{'}) = \sum_{k=1}^{N} \frac{\Sigma_k}{2\pi \sigma_{k}^2 q_{k}^{'}} \exp\left[-\frac{1}{2 \sigma_{k}^2} \left(x^{'2} + \frac{y^{'2}}{q_{k}^{'2}}\right)\right],
    \end{equation}
    where $\Sigma_k$, $\sigma_{k}$, and $q_{k}^{'}$ represent \textcolor{black}{the total stellar mass}, dispersion along the major axis, and axial ratio of the $k$-th Gaussian component, respectively. \textcolor{black}{MGEFIT} outputs the two-dimensional (2D) half-mass radius $R_e$ and semi-major axis $R_e^{\mathrm{maj}}$ \textcolor{black}{based on the 2D stellar mass distribution in the MGE formalism}. It should be noted that the MGE approach does not extrapolate the light/density of a galaxy to infinite radii when determining $R_e$. Beyond three times the dispersion of the largest MGE Gaussian component, the model's flux effectively drops to zero. As a result, this method yields a slightly smaller $R_e$ compared to other techniques \citep{1948AnAp...11..247D, 1987ApJS...64..601B, 1991rc3..book.....D}. The $R_e$ and $R_{e}^{\text{maj}}$ values we use in this paper are thus scaled by a factor of 1.35, as suggested by \citet{2013MNRAS.432.1709C}, which subtracts potential systematic errors due to different methods. \par
    
    We compare the $R_e$ and $M_*$ distributions of TNG galaxies with ATLAS$^{\mathrm{3D}}$ and MaNGA samples in Figure \ref{fig:mass_size_relation}. TNG galaxies \textcolor{black}{(dashed profiles)} agree well with observed \textcolor{black}{(solid profiles)} $R_e$ and $M_*$ values. Those with $M_* \geq 10^{10.8} M_\odot$ may be slightly larger ($\sim2$ kpc), but this has minimal impact on our results. \textcolor{black}{Moreover, we also examine the measurement using $r-$band luminosity images. It reveals that $R_e$ measured by $r$-band fitting \textcolor{black}{yields a slightly larger value} by $\lesssim 1$ kpc than those of mass-based measurements, propagating to a marginal increase of $\Delta\lambda_R(R_e) \lesssim 0.03$. Though quantifiable, these differences negligibly impact our kinematic conclusions.}

\subsubsection{Kinematic maps and their Fourier expansion models}
    
    We compute the line-of-sight mean velocity and velocity dispersion fields for TNG galaxies from a randomly oriented perspective. All images are binned into 0.2 kpc intervals, equivalent to 2 arcseconds for a galaxy at a distance of 20 Mpc. This creates mock kinematic observations comparable to the ATLAS$^{\text{3D}}$ survey \citep{2011MNRAS.413..813C}. Examples of fast (left panels) and slow (right panels) rotators are shown in Figure \ref{fig:FRs_SRs_kinemetry_model}. Stellar particles are grouped into a regular grid centered on the galaxy, spanning a width of 40 kpc. Voronoi binning\footnote{https://pypi.org/project/vorbin/} \citep{2003MNRAS.342..345C} is then applied to merge bins with an insufficient number of stellar particles. Due to the Poissonian noise in particle counts, the target S/N relates to the number of particles per bin as S/N $= \sqrt{N_p}$. To achieve robust kinematics, we set S/N$\geq 10$, requiring at least 100 stellar particles per bin. The mean velocity and velocity dispersion of the $i$-th bin are calculated as:  
    \begin{equation}
    	\begin{split}
    	\overline{V_i} =  \frac{\sum_{n}M_{n,i}V_{n,i}}{\sum_{n}M_{n,i}} ;\quad   
            \sigma_i = \sqrt{\frac{ \sum_{n}M_{n,i}(V_{n,i}-\overline{V_i})^{2}}{\frac{N_i}{N_i -1}\sum_{n}M_{n,i}}}
    	\end{split}
    \end{equation}
    where the index $n$ runs over the stellar particles with mass $M_{n,i}$ within $i$th Voronoi bin, $N_i$ is the particle number of this bin. \par

    We further apply the KINEMETRY package for modeling the velocity maps through a truncated Fourier expansion along elliptical contours \citep{2006MNRAS.366..787K, 2008MNRAS.390...93K,2016MNRAS.457..147F}:
    \begin{equation}
        V(\psi) = V_0 + \sum_{n=1}^{k} \left[A_n\sin(n\psi) + B_n\cos(n\psi)\right]\label{kinemetry_eq}
    \end{equation}
    where $\psi$ denotes the eccentric anomaly, $V_0$ represents the systemic velocity, and $A_n$, $B_n$ are the Fourier coefficients. In the case of an ideal rotating disk, symmetry requirements force all coefficients except $B_1$ to vanish, simplifying the velocity profile to $V(\psi) = V_0 + B_1\cos(\psi)$. The cosine term captures the characteristic signature of regular rotation $B_1 = V_{\text{rot}}$ (the second row of Figure \ref{fig:FRs_SRs_kinemetry_model}), \textcolor{black}{where $V_{\rm rot}$ denotes the velocity amplitude}, while non-zero higher-order coefficients reveal departures from axisymmetry in the velocity field (the third row of Figure \ref{fig:FRs_SRs_kinemetry_model}). \par

 \begin{figure*}[t]
        \centering        \includegraphics[width=0.98\textwidth]{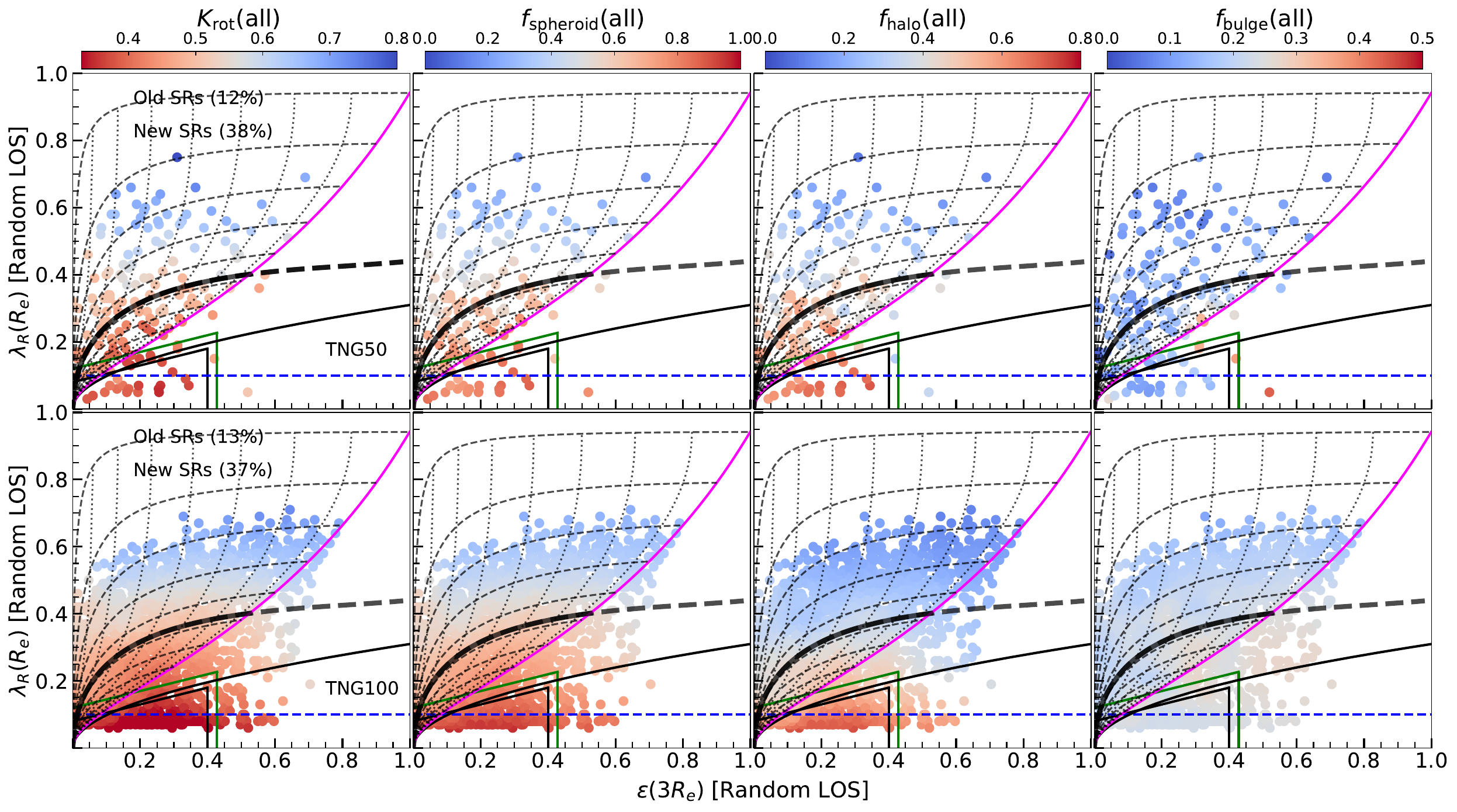}
            \caption{The $\lambda_{R}(R_e)-\varepsilon(3R_e)$ diagram color coded by $\kappa_{\rm rot}(\rm{all})$, $f_{\rm spheroid}(\rm{all})$, $f_{\rm halo}(\rm{all})$, and $f_{\rm{bulge}}(\rm all)$ for TNG50 (upper) and TNG100 (lower). \textcolor{black}{Here, 
            $\kappa_{\rm rot}(\rm all)$, $f_{\rm spheroid}(\rm all)$, 
            $f_{\rm halo}(\rm all)$, and 
            $f_{\rm bulge}(\rm all)$
            represent the significance of cylindrical rotation, the spheroidal mass fraction, the stellar halo mass fraction, and the bulge mass fraction, respectively (see Section \ref{sec_structure_mass_ratio} and \ref{sec_krot})}. The magenta curve shows the $\lambda_{R}-\varepsilon$ relation in edge-on view for axisymmetric galaxies derived by \citet{2007MNRAS.379..418C}. The black dotted curves illustrate this relation across varying $i$, with steps of $\Delta i = 10^{\circ}$. Additionally, \textcolor{black}{the thin black} dashed lines represent the theoretical distribution for different rotations spaced at intervals of $\Delta \lambda_{{R,\rm intr}} = 0.1$. The bold black curve represents our proposed new threshold for distinguishing between SRs and FRs. \textcolor{black}{This threshold is derived from the tensor virial theorem for oblate galaxies with intrinsic parameters \(\lambda_{{R,\rm intr}}(R_e) = 0.4\) and \(\varepsilon_{\text{intr}} = 0.525\), assuming an anisotropy of \(\delta = 0.7\varepsilon_{\text{intr}} = 0.367\)}. For comparison, \textcolor{black}{the thin solid black} line is the standard empirical threshold defined by $\lambda_{R}(R_e) = 0.31 \times \sqrt{\varepsilon}$ proposed in \citet[][eq. 3]{2011MNRAS.414..888E}. We present the proportion of SRs classified by our new threshold alongside the standard old threshold at the upper-left corner. We smooth the color data using the locally weighted regression method LOESS \citep{Cleveland01091988, 2013MNRAS.432.1862C} with a smoothing factor of $0.1$.   \textcolor{black}{The blue horizontal dashed line represents the empirical threshold defined by $\lambda_{R}(R_e) = 0.1$ according to \cite{2007MNRAS.379..401E}, the black polygon denotes the empirical threshold set by $\lambda_{R}(R_e) < 0.08 + \varepsilon/4 \,\, {\rm and} \,\, \varepsilon < 0.4$ as per \cite{2016ARA&A..54..597C}, and the green polygon signifies the empirical threshold given by $\lambda_{R}(R_e) < \lambda_{R\text{start}} + \varepsilon/4 \,\, {\rm and} \,\, \varepsilon < 0.35 + \frac{\lambda_{R\text{start}}}{1.538}$ (with $\lambda_{R\text{start}} = 0.12$ when the IFS data quality is comparable to that  \textcolor{black}{of SAMI}) based on \cite{2021MNRAS.505.3078V}.}}
            \label{fig:TNG_intrinsic_dynamical_auto_GMM}
    \end{figure*}
\begin{figure}[t]
    \centering        
    \includegraphics[width=0.99\columnwidth]{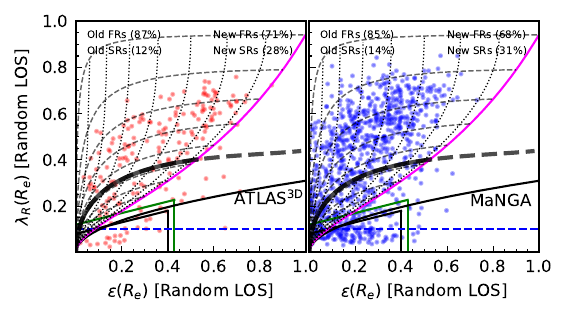}
    \caption{\textcolor{black}{The $\lambda_{\rm R}(R_e)-\varepsilon(R_e)$ diagram of the sample of ATLAS$^{\mathrm{3D}}$(left panel) and MaNGA (right panel).} The proportions of SRs and FRs classified using our new threshold, compared to the standard old definition, are provided at the top. \textcolor{black}{The definitions of all the lines are the same as those in Figure \ref{fig:TNG_intrinsic_dynamical_auto_GMM}.}}
    \label{fig:vsigmae}
\end{figure} 

\section{Parameters quantifying rotation properties of galaxies}\label{KinematicParameter}   

    The relative importance of rotation is a key factor that determines the morphology, assembly history, and internal dynamics of galaxies. Several dimensionless parameters are commonly employed to characterize the rotational properties of galaxies, such as \textcolor{black}{$\lambda_{R}$} (the classical spin parameter), $\kappa_{\text{rot}}$ (cylindrical rotation energy), and structural mass ratios derived from kinematic decomposition. \textcolor{black}{Meanwhile, the higher-order term $\overline{k_5}$ from the Fourier decomposition of velocity fields, which is used to describe kinematic anomalies,} has been used to classify fast and slow rotators in ETGs \citep{2004MNRAS.352..721E, 2011MNRAS.414..888E}, despite lacking a well-defined physical connection to the relative importance of rotation. \textcolor{black}{ $\overline{k_5}$ is defined as the mass-weighted average of the fifth-order kinematic Fourier coefficient $k_5 = \sqrt{A_5^2 + B_5^2} / V_{\text{rot}}$ within the effective radius $R_e$, \textcolor{black}{ where $V_{\rm rot}$ represents the velocity amplitude}, $A_5$ and $B_5$ are the fifth sine and cosine terms, respectively, in the truncated Fourier expansion for modeling the velocity field along elliptical contours (see Equation \ref{kinemetry_eq})}. The relationships among these parameters and their consistency with observational data are not yet fully understood.\par

    This study systematically estimates and compares these parameters through: (1) rigorous evaluation of their diagnostic capabilities, (2) quantitative analysis of their inter-parameter correlations, and (3) direct comparison with observational results. This comprehensive approach seeks to establish a universal framework for characterizing galactic rotation and classifying of slow and fast rotators of galaxies in both observations and simulations.\par

    \begin{figure*}[t]
    \centering
    \includegraphics[width=\textwidth]{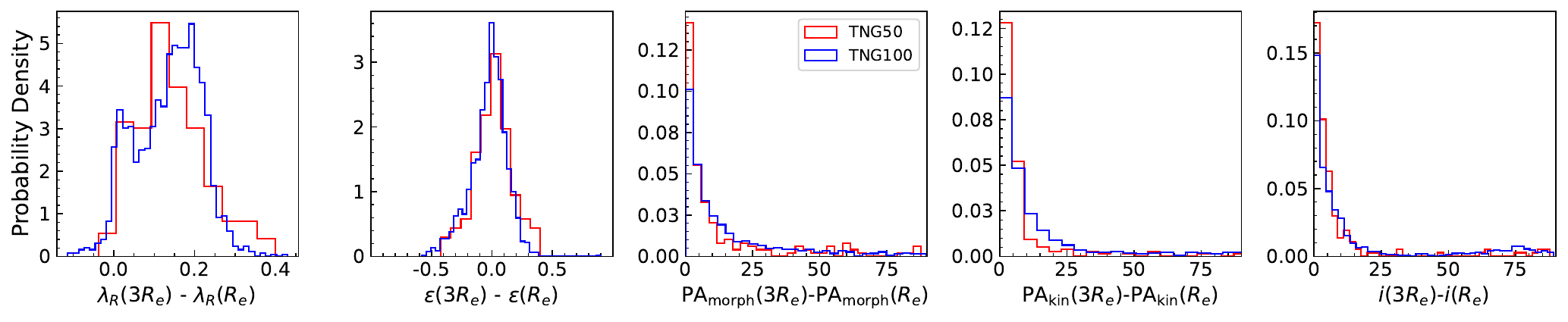}
        \caption{ 
        \textcolor{black}{The distributions of the differences between various galactic parameters measured within (at) $3R_e$ and $R_e$ iso-density ellipses. From left to right, it depict various galactic parameters, specifically: $\lambda_{R}(3R_e)-\lambda_{R}(R_e)$, $\varepsilon(3R_e) - \varepsilon(R_e)$,  $\text{PA}_{\text{morph}}(3R_e) -\text{PA}_{\text{morph}}(R_e)$,  $\text{PA}_{\text{kin}}(3R_e) - \text{PA}_{\text{kin}}(R_e)$, and $i(3R_e)-i(R_e)$. Here, \textcolor{black}{$\lambda_{R}(3R_e)$}, $\varepsilon(3R_e)$,
        $\text{PA}_{\text{morph}}(3R_e)$, $\text{PA}_{\text{kin}}(3R_e)$, $i(3R_e)$ represent the respective parameters measured within (at) $3R_e$, while \textcolor{black}{$\lambda_{R}(R_e)$}, $\varepsilon(R_e)$,
        $\text{PA}_{\text{morph}}(R_e)$, $\text{PA}_{\text{kin}}(R_e)$, $i(R_e)$ denote the same parameters but measured within (at) $R_e$ isophoto. The red and blue histograms signify the galaxies in TNG50 and TNG100, respectively. }        
        }
    	\label{fig:parameter_1re_2re}
    \end{figure*}

 \subsection{The \texorpdfstring{\textcolor{black}{$\lambda_{R}$}-$\varepsilon$}{} diagram}\label{sec_lambda_epsilon}
   The spin parameter \textcolor{black}{$\lambda_{R}$} is a dimensionless parameter that quantifies the rotation of galaxies as defined by \cite{2011MNRAS.414..888E}
    \begin{equation}
    	\begin{split}
    	\lambda_R  \equiv \frac{\langle R \mid \overline{V} \rangle}{\langle R  \sqrt{\overline{V}^{2} + \sigma^{2}} \rangle}  = \frac{ \sum_{i} M_{i} R_{i} |\overline{V_{i}}|}{ \sum_{i} M_i R_{i} \sqrt{\overline{V_i}^{2} + \sigma_{i}^{2}}  }
    	\end{split}
    \end{equation}        
    where the weighting with the flux is substituted here with a weighting with \textcolor{black}{the mass contained within the $i$th Voronoi bin, $M_i$.} \textcolor{black}{$R_{i}$ is the distance of the $i$th bin to the galaxy center.} \textcolor{black}{$\lambda_{R}(R_e)$} and \textcolor{black}{$\lambda_{R}(3R_e)$} are computed by summing over all Voronoi bins \textcolor{black}{within ellipses of one and three times $R_e$, with semi-major axis $R_e^{\rm maj}$ \citep{2018MNRAS.477.4711G}\footnote{https://github.com/marktgraham/lambdaR\_e\_calc}.}\par

    Galaxies with higher rotational velocities typically exhibit larger spin parameters (\textcolor{black}{$\lambda_{R}$}) and lower ellipticities ($\varepsilon$), as illustrated in the \textcolor{black}{$\lambda_{R}$}-$\varepsilon$ diagrams of Figures \ref{fig:TNG_intrinsic_dynamical_auto_GMM} and \ref{fig:vsigmae} for both TNG simulations and IFS observations. \textcolor{black}{
    Our analysis confirms a substantial galaxy population in the high-ellipticity region (\(\varepsilon \gtrsim 0.4\)) with intermediate-to-low \(\lambda_{R}(R_e)\) values—occupying the right portion of the \(\lambda_{R}(R_e)\)–\(\varepsilon(R_e)\) plane. \citet{2020A&A...641A..60P} argued that such galaxies (\(\sim\)50\% of total ETG samples) should be excluded as non-physical via applying an axis-ratio cut at \(R_e\) as suggested by observations \citep{2014MNRAS.444.3340W,2017MNRAS.472..966F,2018ApJ...863L..19L,2018MNRAS.479.2810E}. Excluding such galaxies may induce severe selection bias. While ellipticity (\(\varepsilon\)) is commonly used to quantify inclination of galaxies, we find that these specific systems are predominantly barred galaxies, whose central \(\varepsilon(R_e)\) are enhanced by the bar structure. There is no robust justification for excluding such galaxies, as the presence of elongated bar structures is unlikely to significantly alter overall galactic kinematics. To mitigate inclination errors from this central triaxiality of bars and the associated bias of excluding such galaxies, we adopt \(\varepsilon\) measured at \(3R_e\).} 17 of 260 galaxies from \textsc{ATLAS$^{\mathrm{3D}}$} are measured at $2.5R_e-3R_e$, and the majority of observational data is measured at $R_e$. This issue should not significantly affect our conclusion for fast-rotating cases. \citet{2011MNRAS.414..888E} introduced an empirical threshold $\lambda_R(R_e) = 0.31 \sqrt{\varepsilon}$ (thin black curve in Figure~\ref{fig:vsigmae}) to distinguish fast rotators from slow rotators. This criterion is initially defined by the fifth-order kinematic Fourier coefficient \textcolor{black}{$\overline{k_5}$}, which quantifies non-axisymmetric rotational components. For instance, the bottom-right panel of Figure~\ref{fig:FRs_SRs_kinemetry_model} showcases a slow rotator with elevated \textcolor{black}{$\overline{k_5}$} values \textcolor{black}{(see details in Section \ref{KinematicParameter})}. This \textcolor{black}{classical} criterion \textcolor{black}{yields} about 12\% SRs in TNG simulations, which is consistent with the result of both \textsc{ATLAS$^{\mathrm{3D}}$} and MaNGA. \par

    To assess the uncertainty in kinematic measurements at different galactocentric radii, we compute the deviation between \textcolor{black}{$\lambda_{R}$} values measured within $R_e$ and $3R_e$, as shown in Figure~\ref{fig:parameter_1re_2re}. The differences between morphologically derived position angles ($\text{PA}_{\text{morph}}$) and kinematically derived ones ($\text{PA}_{\text{kin}}$) are generally minor (\textcolor{black}{By applying KINEMETRY to perform elliptical fitting on simulated galaxy images, the $\text{PA}_{\rm morph}$ and $\text{PA}_{\rm phot}$ can be derived.}), i.e., $<20$ degree. In contrast, \textcolor{black}{$\lambda_{R}(R_e)$} values are systematically lower than \textcolor{black}{$\lambda_{R}(3R_e)$}, which is expected due to the kinematic suppression often caused by central slow-rotating components, e.g, bulges. This radial dependence highlights the importance of well-defined apertures in kinematic analyses, particularly for galaxies with structurally complex components. For consistency, we adopt $3R_e$ measurements to represent the \textcolor{black}{accurate} galactic properties, while $R_e$ measurements are retained for direct comparison with observational studies.
    
    \textcolor{black}{In addition, the inclinations for the observational data are derived from the JAM modeling, which is quite time-consuming. Therefore, we then use the simple correlation $i = \rm{arccos} \sqrt{1-\varepsilon}$ that allows us to apply our method to more IFS data where dynamical modeling is not available. we compute the deviation between $i$ values measured within $R_e$ and $3R_e$, as shown in the right-most panel of Figure~\ref{fig:parameter_1re_2re}. The differences between $i(3R_e)$ and $i(R_e)$ are generally minor}.

\subsection{Approximation of the intrinsic spin parameter      \texorpdfstring{\textcolor{black}{$\lambda_{R,\rm{intr}}$}}{}}\label{lambda_intr_label}

    The location of galaxies within the \textcolor{black}{$\lambda_{R}$}-$\varepsilon$ diagram is influenced not only by their rotational properties but also by their inclination angles. To correct for this projection effect, we estimate an intrinsic spin parameter, \textcolor{black}{$\lambda_{R,\rm{intr}}$}, which approximates the spin value for an edge-on configuration ($i = 90^\circ$), under the assumption that galaxies are axisymmetric systems. Following the method of \cite{1987gady.book.....B}, we derive the edge-on $(V/\sigma)$ as:
    \begin{equation} \label{eq_v_sigma_e_intr}
    \left(\frac{V}{\sigma}\right)_{\text{e}} = \left(\frac{V}{\sigma}\right)_{\text{obs}} \frac{\sqrt{1 - \delta \cos^{2} i}}{\sin i}
    \end{equation}
    where the observed ratio $(\frac{V}{\sigma})_{\text{obs}} = \sqrt{\frac{\sum_j M_j \overline{V_j}^2}{\sum_j M_j \sigma_j^2}}$ is computed along the line-of-sight for a given inclination angle $i$. $\delta$
    is the velocity anisotropy that is randomly sampled from a uniform
    distribution in the empirically-motivated range $[0, 0.7\varepsilon_{\rm intr}]$, as suggested by \citet{2007MNRAS.379..418C}. We derive the edge-on
    ellipticity by $\varepsilon_{\rm intr} = 1 - \sqrt{1 + \varepsilon(\varepsilon -2)/\text{sin}^2 \,i}$. We then approximate the intrinsic spin parameter \textcolor{black}{$\lambda_{R,\rm{intr}}$} by the empirical formula from \cite{2007MNRAS.379..401E, 2011MNRAS.414..888E}
        \begin{equation} \label{eq_lambda_re_intr}
    \lambda_{R,\rm intr} \approx \frac{a(\frac{V}{\sigma})_{\text{e}}}{\sqrt{1+a^2(\frac{V}{\sigma})_{\text{e}}^{2}}}
    \end{equation}
    where $a=1.1$. \textcolor{black}{$\lambda_{R,\rm{intr}}$} we derive roughly corresponds to the values of \textcolor{black}{$\lambda_{R}$} projected onto the magenta curve in Figures \ref{fig:TNG_intrinsic_dynamical_auto_GMM} and \ref{fig:vsigmae}.
    
\begin{figure}[t]
        \centering        \includegraphics[width=0.98\columnwidth]{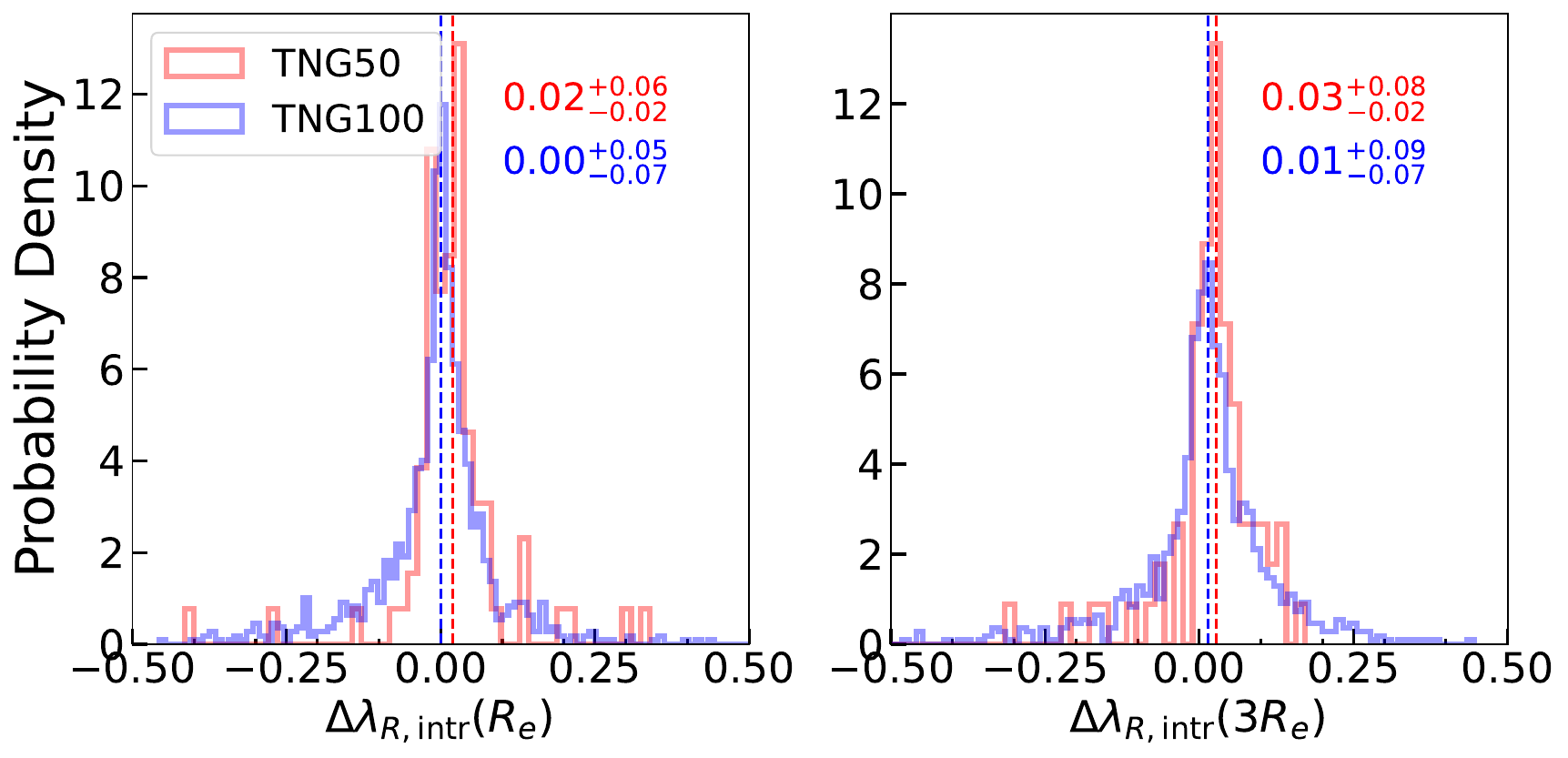}
            \caption{The distribution of the deviation \textcolor{black}{$\Delta \lambda_{R,\rm intr}$} within different regions between the intrinsic spin parameter we approximate and its edge-on projection. \textit{Left panel}: the distribution of the deviation \textcolor{black}{$\Delta \lambda_{R,\rm intr}(R_e)$}. \textit{Right panel}: the distribution of the deviation \textcolor{black}{$\Delta \lambda_{R, \rm intr}(3R_e)$}. The red histogram signifies the galaxy sample derived from TNG50, whereas the blue histogram represents the galaxy sample from TNG100. The vertical dashed lines show the median of the intrinsic spin spectroscopic parameter estimation error. \textcolor{black}{The median and the 16th and 84th percentiles are listed on the top right of each subplot.}} 
        \label{fig:TNG_intrinsic_lambdare_vs_theory_lambdre}
    \end{figure}

    To assess how accurately this method recovers \textcolor{black}{$\lambda_{R,\rm{intr}}$}, we calculate the deviation $\Delta \lambda_{\rm{R,intr}}$ between the intrinsic spin parameter we approximate and its real values in edge-on views, \textcolor{black}{specifically for the TNG galaxies}, as shown in Figure~\ref{fig:TNG_intrinsic_lambdare_vs_theory_lambdre}. The distributions for TNG50 (red histograms) and TNG100 (blue histograms) reveal no significant differences in \textcolor{black}{$\lambda_{R,\rm{intr}}$} measurements within $R_e$ (left panel) versus within $3R_e$ (right panel). The results demonstrate that the method reliably recovers the intrinsic spin parameter, with deviations falling within acceptable limits. 

\subsection{Mass fractions of kinematically-derived structures using \texttt{auto-GMM}: spheroid \texorpdfstring{$f_{\rm spheroid}$}{} and stellar halo \texorpdfstring{$f_{\rm halo}$}{}}\label{sec_structure_mass_ratio}

   \begin{figure*}[t]
        \centering        \includegraphics[width=0.98\textwidth]{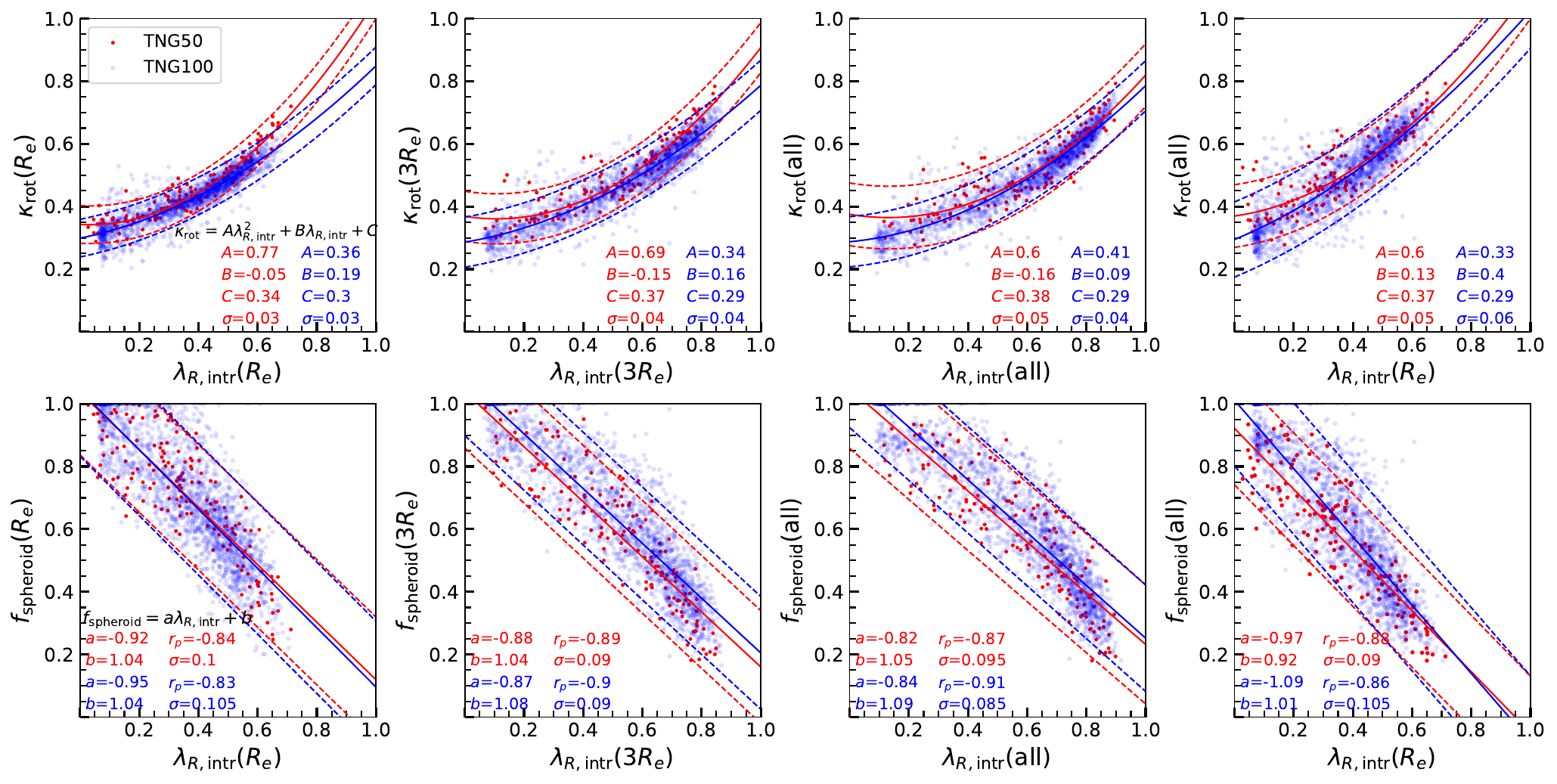}
            \caption{The distribution and correlation of $\kappa_{\text{rot}}$ vs. \textcolor{black}{$\lambda_{R,\rm{intr}}$} (upper) and $f_{\rm spheroid}$ vs. \textcolor{black}{$\lambda_{ R,\rm intr}$} (lower) measured within different regions from left to right. The solid red and blue curves show the best-fit results for the TNG50 and TNG100 datasets, respectively, while the dashed curves represent their $2\sigma$ \textcolor{black}{confidence interval}. The corresponding fitting coefficients and $\sigma$ values are listed in the bottom-right corner of each panel. In the lower panels, we additionally provide the Pearson correlation coefficient ($r_p$). The value of $r_p\sim-1$ indicates a strong anti-correlation between \textcolor{black}{$\lambda_{R,\rm{intr}}$} and $f_{\rm spheroid}$.
            } 
        \label{fig:TNG_intrinsic_lambdare_vs_dynamical_3Re_krot_spheroid}
    \end{figure*}

    Recent works \citep{2019ApJ...884..129D, 2020ApJ...895..139D, 2021ApJ...919..135D} introduced \texttt{auto-GMM}, an automated unsupervised method for accurately decomposing galaxy kinematic structures in simulation data. This approach analyzes stellar particles in a three-dimensional kinematic phase space comprising the circularity parameter $\epsilon = j_z/j_c(e)$ \citep{2003ApJ...597...21A}, the non-azimuthal angular momentum ratio $j_p/j_c(e)$, and the normalized binding energy $e/|e|_{\text{max}}$ \citep{2012MNRAS.421.2510D}. $j_z$ denotes the specific angular momentum component aligned with the galaxy's minor axis ($z$-axis) in cylindrical rotation. This quantity is normalized by $j_c(e)$, the maximum possible circular angular momentum corresponding to the particle's binding energy $e$. The parameter $j_z/j_c(e)$ thus quantifies rotation aligned with the galaxy's net angular momentum, whereas $j_p/j_c(e)$ measures misaligned rotation, and $e/|e|_{\text{max}}$ indicates how tightly bound each stellar particle is.  

    When applied to TNG simulations, this method successfully identified distinct structural components, including cold disk, warm disk, bulge, and stellar halo. The total spheroidal mass fraction $f_{\text{spheroid}}$ is obtained by combining the contributions from both the bulge ($f_{\text{bulge}}$) and the stellar halo ($f_{\text{halo}}$), with data taken from \citet{2020ApJ...895..139D} and \citet{2021ApJ...919..135D} \footnote{https://www.tng-project.org/data/docs/specifications/\#sec5}. The measurements are provided at three different scales: $f(R_e)$ for stars within 1$R_e$, $f(3R_e)$ within 3$R_e$, and $f({\rm all})$ for all stars in each galaxy. $f_{\text{halo}}$ defined in this way correlates closely with the strength of merger history \citep{2021ApJ...919..135D, Proctor2024}. One of the key advantages of \texttt{auto-GMM} is its allowance for weakly rotating spheroids, differing from other studies \citep{2022MNRAS.515.1524Z, 2024A&A...692A..63C, 2025MNRAS.541.2304L} which assume zero rotation in bulges and stellar halos—a likely unrealistic assumption.

    The spheroidal mass fraction $f_{\text{spheroid}}$ serves as a quantitative measure for determining the relative dominance of spheroidal components in galaxies. Based on this metric, we can classify galaxies with $f_{\text{spheroid}}<0.5$ as FRs which means disk structures dominate them, while those with $f_{\text{spheroid}}>0.5$ are identified as SRs.

    \subsection{Assessing the significance of cylindrical rotation energy \texorpdfstring{$\kappa_{\text{rot}}$}{}}\label{sec_krot}
   
    The significance of cylindrical rotation is quantified by the parameter \textcolor{black}{$\kappa_{\text{rot}} = \frac{K_{\text{rot}}}{K} = \frac{1}{K}\sum\left[\frac{1}{2}m_{i}\left(\frac{j_{z,i}}{R_{i}}\right)^2\right]$},
    as defined in \citep{2010MNRAS.409.1541S}, where \textcolor{black}{$K$ and $m$ represent} the total kinetic energy and mass of each stellar particle, respectively. This widely-used metric provides a clear physical interpretation of ordered rotation in simulated galaxies and is computationally straightforward to evaluate.\par 

    However, several important considerations should be noted. First, while commonly adopted, the threshold of $\kappa_{\text{rot}} = 0.5$ for distinguishing between disk galaxies (corresponding to FRs) and elliptical galaxies (SRs) lacks strong theoretical justification \textcolor{black}{\citep{Zhao_2020,2021MNRAS.508..926R,2025A&A...697A.236L}}. Second, for a completely dispersion-dominated system, theoretical considerations predict $\kappa_{\text{rot}} = 1/3$. Finally, the physical interpretation remains ambiguous for both disk-dominated galaxies ($\kappa_{\text{rot}} > 0.5$) and elliptical galaxies ($\kappa_{\text{rot}} < 0.5$). These limitations highlight the need for careful interpretation when applying $\kappa_{\text{rot}}$ criteria to different galactic systems.\par 

\section{Approximating \texorpdfstring{$\kappa_{\text{rot}}$}{} and \texorpdfstring{$f_{\rm spheroid}$}{} from scaling relations with \texorpdfstring{\textcolor{black}{$\lambda_{R,\rm{intr}}$}}{}}\label{label_scaling_krot_spheroid}

    The intrinsic spin parameter \textcolor{black}{$\lambda_{R,\rm{intr}}$}, together with the $\lambda$-$\varepsilon$ diagram, offers a relatively accessible observational metric. A higher \textcolor{black}{$\lambda_{R,\rm{intr}}$} value signifies that a galaxy's kinematic structure is dominated by ordered rotation, while a lower value indicates a greater influence of random motion. Consequently, Figure \ref{fig:TNG_intrinsic_dynamical_auto_GMM} reveals a clear trend: both $f_{\text{spheroid}}$ and $\kappa_{\text{rot}}$ decrease as \textcolor{black}{$\lambda_{R,\rm{intr}}$} declines. In this section, we then quantitatively examine the scaling relations connecting \textcolor{black}{$\lambda_{R,\rm{intr}}$}, $\kappa_{\text{rot}}$, and $f_{\text{spheroid}}$.

    We begin by investigating the relationship between \textcolor{black}{$\lambda_{R,\rm{intr}}$} and $\kappa_{\text{rot}}$. The fundamental difference in their velocity dependence — linear for \textcolor{black}{$\lambda_{R,\rm{intr}}$} versus quadratic for $\kappa_{\text{rot}}$ — motivates a quadratic fitting function of the form:
        \begin{equation}
        \kappa_{\text{rot}} = A\lambda_{{R,\rm intr}}^2 + B\lambda_{{R,\rm intr}} + C
        \end{equation}
    where $A$, $B$, and $C$ represent the coefficients to be determined through regression analysis. This functional form naturally accommodates their distinct velocity scalings while enabling quantitative comparison of their kinematic relationship. The first row of Figure~\ref{fig:TNG_intrinsic_lambdare_vs_dynamical_3Re_krot_spheroid} reveals a robust correlation between \textcolor{black}{$\lambda_{R,\rm{intr}}$} and $\kappa_{\text{rot}}$ \textcolor{black}{(see the fitting result at the bottom-right corners)}. This relationship holds consistently across both TNG50 (red) and TNG100 (blue) simulations, regardless of the measurement aperture - whether considering stars within $R_e$, $3R_e$, or the entire system. Notably, as demonstrated in the right-most panel, \textcolor{black}{$\lambda_{{R,\rm intr}}(R_e)$} alone can be used to approximate the global $\kappa_{\text{rot}}(\text{all})$ with an acceptable uncertainty of $\sim 0.05$, despite being measured only within the effective radius ($R_e$). Furthermore, the criterion $\kappa_{\rm rot} > 0.5$ has become a standard threshold for identifying disk-dominated systems in numerical simulations, approximately equivalent to $\lambda_{R,\rm intr}(R_e) \gtrsim 0.4$.\par 
      
    Our analysis further reveals a linear scaling relation between \textcolor{black}{$\lambda_{R,\rm{intr}}$} and $f_{\text{spheroid}}$, as shown in the lower panels of Figure~\ref{fig:TNG_intrinsic_lambdare_vs_dynamical_3Re_krot_spheroid}. This correlation robustly validates \textcolor{black}{$\lambda_{R,\rm{intr}}$} as an effective diagnostic for quantifying spheroid mass fractions across different measurement apertures, see the fitting result at the bottom-right corners. The observed anti-correlation follows $f_{\rm spheroid} \approx 1 - \lambda_{{R,\rm intr}}$ (or equivalently, $f_{\rm disk} \approx \lambda_{{R,\rm intr}}$), indicating that \textcolor{black}{$\lambda_{R,\rm{intr}}$} directly traces disk dominance with a typical uncertainty of $\sigma \sim 0.1$. This establishes \textcolor{black}{$\lambda_{R,\rm{intr}}$} as a powerful observational proxy for connecting IFS kinematics to the intrinsic structural decomposition of galaxies. Even when limited to one effective radius (\textcolor{black}{$\lambda_{{R,\rm intr}}(R_e)$}) - the typical spatial coverage of IFS observations - \textcolor{black}{$\lambda_{R,\rm{intr}}$} retains diagnostic power for characterizing global galaxy rotation.

     \begin{figure}[t]
        \centering        \includegraphics[width=0.99\columnwidth]{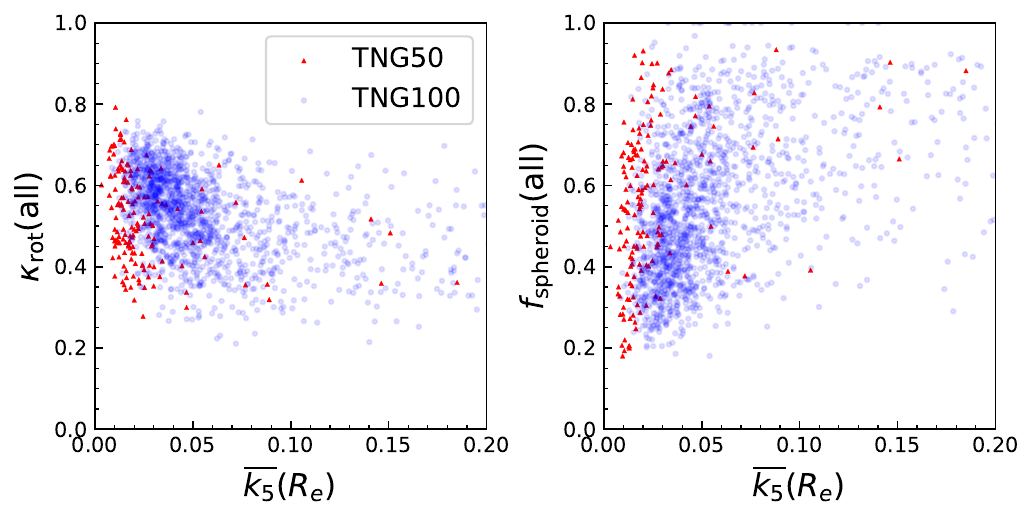}
            \caption{\textcolor{black}{ $\kappa_{\rm rot}(\text{all})$, $f_{\rm spheroid}(\text{all})$ of the selected ETGs samples in TNG simulation, plotted as a function of $\overline{k_{5}}(R_e)$}. No clear correlation between $\kappa_{\text{rot}}(\text{all})$ and $f_{\text{spheroid}}(\text{all})$ of the selected ETGs samples in the TNG simulations. Red and blue dots represent \textcolor{black}{the samples of TNG50 and TNG100}, respectively.} 
        \label{fig:TNG_k5B1_intrinsic_kinematic}
    \end{figure}

    The parameter \textcolor{black}{$\lambda_{R}$} thus can serve as a proxy for estimating \textcolor{black}{$\lambda_{R,\rm{intr}}$}, $\kappa_{\rm rot}$, and $f_{\rm spheroid}$. Empirically, \textcolor{black}{$\lambda_{{R,\rm intr}}(R_e)$} provides a reasonable approximation of the global rotational properties, specifically $\kappa_{\text{rot}}(\text{all})$ and $f_{\text{spheroid}}(\text{all})$. Moreover, it is worth highlighting that the \textcolor{black}{$\lambda_{R,\rm{intr}}$}-$f_\text{spheroids}$ relationship is primarily determined by $f_{\rm halo}$, shown in the third column of Figure \ref{fig:TNG_intrinsic_dynamical_auto_GMM}. Bulges only contribute to scatter without altering the fundamental scaling relation, as shown in the fourth column. \textcolor{black}{In the fourth column}, we further leverage the empirical correlations between \textcolor{black}{$\lambda_{{R,\rm intr}}(R_e)$} and $\kappa_{\text{rot}}(\text{all})$, $f_{\text{spheroid}}(\text{all})$ derived from TNG50 simulations to infer approximate values of $\kappa_{\text{rot}}(\text{all})$ and $f_{\text{spheroid}}(\text{all})$ based on \textcolor{black}{$\lambda_{{R,\rm intr}}(R_e)$} \footnote{\textcolor{black}{The catalog is publicly available.}} \textcolor{black}{for observed galaxies}. \textcolor{black}{A subset of MaNGA galaxies, specifically nine randomly selected sample galaxies, has their detailed information listed in Table \ref{tab:APPENDIX_lambda_intr_krot_sph_obs}. The complete content of MaNGA, $\text{ATLAS}^{\text{3D}}$, TNG50, and TNG100 data is publicly available.}

\section{Kinematic classification of galaxies into fast (disk) and slow (elliptical) rotating galaxies}\label{Kinematic classification}

    \textcolor{black}{In this study, we equate disk galaxies with FRs and elliptical galaxies with SRs. Regarding this, it should be noted that morphological classifications (disk vs. elliptical galaxies) and kinematic classifications (FRs vs. SRs) describe fundamentally distinct galaxy properties. Disks and ellipticals denote visual structures observed photometrically \citep[e.g., spiral arms vs. \textcolor{black}{smooth ellipsoids;}][]{1926ApJ....64..321H,1936rene.book.....H,2005ARA&A..43..581S}, while FRs and SRs categorize galaxies based on dynamical properties quantified by spin parameter $\lambda_{R}$ \citep{2007MNRAS.379..401E,2016ARA&A..54..597C}. Historically, these classifications correlated strongly—ellipticals were presumed pressure-supported SRs, while disks were rotationally dominated FRs—driving early unification in literature. However, IFS surveys like ATLAS$^{\rm 3D}$, SAMI, CALIFA and MaNGA have revealed significant exceptions: part of morphological ellipticals exhibit FR kinematics \citep{2011MNRAS.414..888E,2015MNRAS.454.2050F}, and a minority of disks show SR-like suppression of rotation \citep{2015IAUS..311...78F,2018MNRAS.477.4711G}. Thus, equating disks/ellipticals directly with FRs/SRs overlooks dynamically complex populations and risks oversimplification.}\par

     \begin{table}[ht]
    	\centering  
        \small
        \setlength{\tabcolsep}{4.5pt}
    	\caption{Table with main properties of 933 MaNGA galaxies used in this paper. }  \label{tab:APPENDIX_lambda_intr_krot_sph_obs}
    	\begin{tabular}{cccccccr}    		
            \hline
    		   Plate-IFU & $i$ & $\varepsilon$ &$\lambda_{R}({R_e})$ & $\lambda_{R,\rm intr}(R_e)$ & $\kappa_{\rm rot}(\text{all})$ & $f_{\rm spheroid}(\text{all})$ \\
               (1) & (2) &(3) & (4)&(5)&(6)&(7) \\
    		\hline 
               8329-1901 &  32$^{\circ}$ &  0.12 &  0.46 & 0.63&0.68&0.30\\
               8319-1901 & 33$^{\circ}$ &  0.07 &  0.46 & 0.66&0.69&0.28\\
              8243-6102 & 32$^{\circ}$ &  0.15 &  0.52 & 0.66&0.70&0.28\\
              8243-3701 & 71$^{\circ}$ &  0.24 &  0.22 & 0.23&0.39&0.69\\
              
              8243-6104 & 38$^{\circ}$ &  0.22 &  0.16 & nan& nan&nan\\

              8243-3703 & 80$^{\circ}$ &  0.01 &  0.49& 0.50&0.57&0.43\\

              8439-3702 & 87$^{\circ}$ &  0.53 &  0.71 & 0.71&0.74&0.22\\

              8326-1901 & 47$^{\circ}$ &  0.18 &  0.57 & 0.67&0.70&0.27\\
             
              8326-3701 & 34$^{\circ}$ &  0.12 &  0.43 & 0.59&0.65&0.34\\
             
             \hline
    	\end{tabular}
        \tablefoot{Column (1): The ID associated with a specific observation for a unique MaNGA galaxy. Column (2): The inclination. Column (3): The ellipticity. Column (4): The stellar spin parameter. Column (5): The intrinsic stellar spin parameter. Column (6): The significance of cylindrical rotation. Column (7) the spheroid mass fraction. For the $\lambda_{R,\rm intr}(R_e)$ showing nan values, it means the galaxy's inclination or ellipticity is inaccurate, or the galaxy fails to meet the axisymmetric assumption, making it impossible to deduce the intrinsic ellipticity from the observed ellipticity and inclination.
        Among the 933 \textcolor{black}{MaNGA galaxies, 97 galaxies} do not have $\lambda_{R,\rm intr}(R_e)$ as nan. Among the 193 \textcolor{black}{$\text{ATLAS}^{\rm 3D}$ galaxies, 15 galaxies} have \textcolor{black}{$\lambda_{R,\rm{intr}}$} as nan. A complete table is available from the journal website.}
        
    \end{table}

    \textcolor{black}{Therefore, this work introduces unified classification thresholds that reconcile galaxy morphology and kinematics (see Section \ref{lab_classity_by_GMM}). Our criteria reclassify approximately 4\% of previously identified disk galaxies as ellipticals and approximately 16\% of fast rotators as slow rotators. With the associated publicly released catalog (Table \ref{tab:APPENDIX_lambda_intr_krot_sph_obs}), we establish a consistent disk/FR and elliptical/SR mapping applicable to both observational surveys and cosmological simulations.}

\subsection{Failure of the empirical classification of slow/fast rotators: the high-order Fourier kinematic components \texorpdfstring{$\overline{k_5}$}{} and the empirical \texorpdfstring{$\lambda_{R}(R_e)-\varepsilon$}{} criterion}
\label{sec:k5}

    To evaluate the reliability of the standard FRs/SRs classification in previous studies \citep[e.g.,][]{2011MNRAS.414.2923K, 2011MNRAS.414..888E}, we also  calculate the \( \overline{k_5} \) (see Section \ref{KinematicParameter}). However, we find that \( \overline{k_5} \) exhibits no significant correlation with either \( \kappa_{\rm rot} \) or \( {f}_{ \rm spheroid } \), suggesting it poorly traces kinematic disk structure or global rotation (Figure \ref{fig:TNG_k5B1_intrinsic_kinematic}), \textcolor{black}{where the underlying assumption here is that galaxies with \( f_{\rm spheroid} < 0.5 \) or \( \kappa_{\rm rot} > 0.5 \) are classified as FRs (disk-dominated), while those with \( f_{\rm spheroid} > 0.5 \) or \( \kappa_{\rm rot} < 0.5 \) are classified as SRs (spheroid-dominated).} This implies that both \( \overline{k_5} \) and the standard \( \textcolor{black}{\lambda_{R}}-\varepsilon \) criterion (thin black profiles in Figures \ref{fig:TNG_intrinsic_dynamical_auto_GMM} and \ref{fig:vsigmae}) inadequately distinguish SRs from FRs, though they successfully identify extreme slow-rotating cases \citep{2023MNRAS.526.1022L}.
    
    \textcolor{black}{Moreover, SRs are typically identified using also other empirically derived thresholds in the \(\lambda_{R}(R_e)-\varepsilon\) plane. Common selection criteria include:
    1) \cite{2007MNRAS.379..401E}'s constant threshold \(\lambda_{R}(R_e) < 0.1\) (blue dashed line in Figure \ref{fig:TNG_intrinsic_dynamical_auto_GMM});
    2) \cite{2011MNRAS.414..888E}'s scaling relation \(\lambda_{R}(R_e) < 0.31\sqrt{\varepsilon}\) (black curve);
    3) \cite{2016ARA&A..54..597C}'s polygon criterion requiring \(\lambda_{R}(R_e) < 0.08 + \varepsilon/4\) with \(\varepsilon < 0.4\) (black polygon); and
    4) \cite{2021MNRAS.505.3078V}'s adaptable threshold \(\lambda_{R}(R_e) < \lambda_{R\text{start}} + \varepsilon/4\) where \(\varepsilon < 0.35 + \lambda_{R\text{start}}/1.538\) (green polygon), adopting \(\lambda_{R\text{start}} = 0.12\) for SAMI-like data quality.
    Given the functional similarity of these thresholds—particularly their convergence near \(\lambda_{R}(R_e) = 0.31\sqrt{\varepsilon}\)—we designate the \cite{2011MNRAS.414..888E} relation as our reference criterion for distinguishing classical fast rotators from slow rotators.}
    A revised threshold, better quantifying galactic rotation, is needed to reliably classify SRs in observations and enable robust comparisons with simulations.\par

     \begin{figure}[t]
        \centering        \includegraphics[width=0.98\columnwidth]{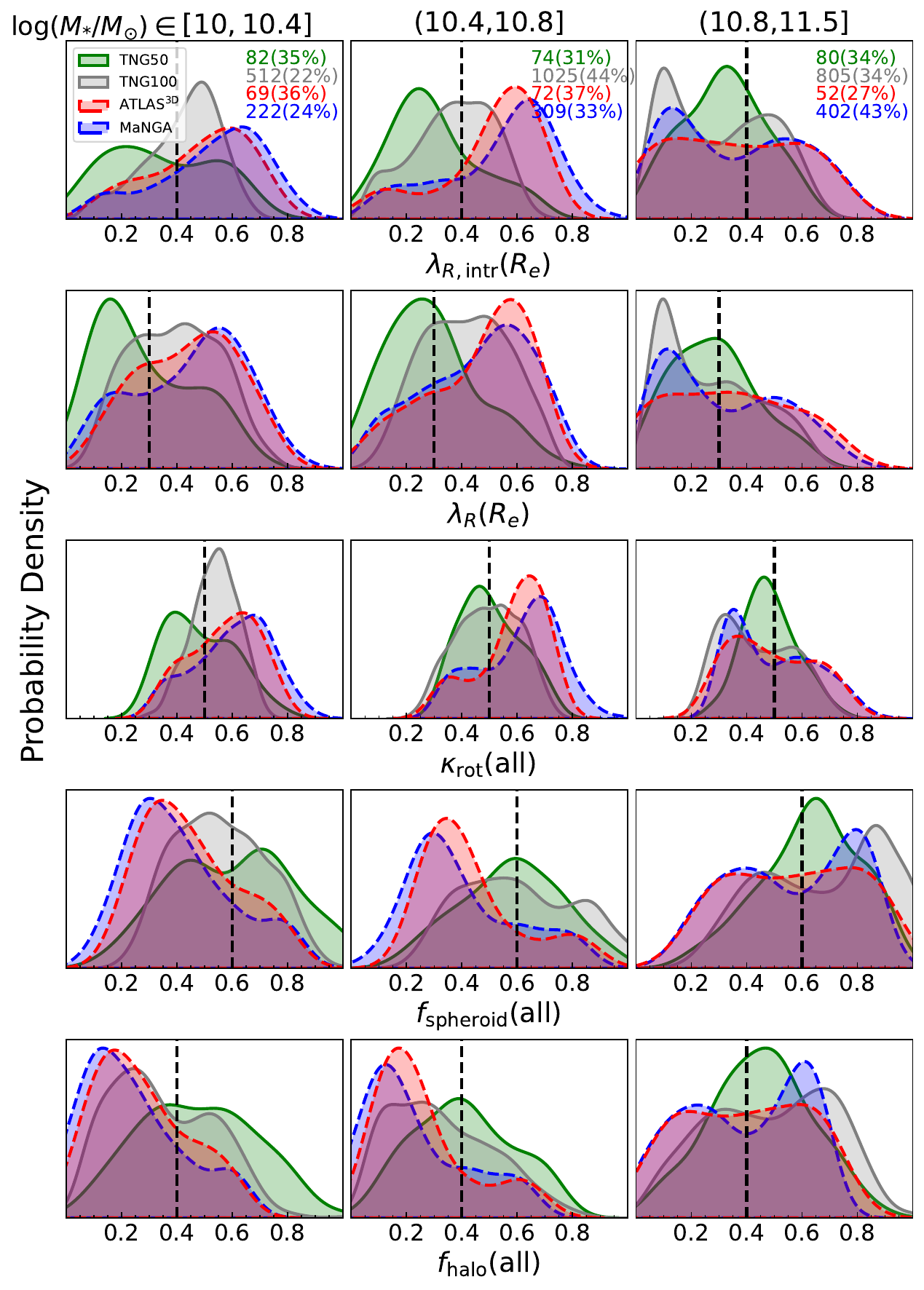}
            \caption{The probability density distributions of \textcolor{black}{$\lambda_{{R,\rm intr}}(R_e)$, $\lambda_{R}(R_e)$}, $\kappa_{\text{rot}}(\text{all})$, $f_{\text{spheroid}}(\text{all})$, and $f_{\text{halo}}(\text{all})$. \textcolor{black}{The green solid, gray solid, blue dashed, and red dashed profiles} represent ETG samples of TNG50, TNG100, MaNGA, and ATLAS$^{\text{3D}}$, respectively. The panels display three stellar mass ranges of $10^{10}-10^{10.4} M_\odot$, $10^{10.4}-10^{10.8} M_\odot$, and $10^{10.8}-10^{11.5} M_\odot$, arranged from left to right. In the \textcolor{black}{top right corner of each upper panel}, we give the proportion and absolute count of galaxies within each mass range relative to the total ETG sample. \textcolor{black}{From top to bottom, \textcolor{black}{these vertical black dashed lines} represent $\lambda_{ R,\rm intr}(R_e) = 0.4$, $\lambda_{R}(R_e) = 0.3$, $\kappa_{\rm rot}(\text{all}) = 0.5$, $f_{\rm spheroid}({\rm all}) = 0.6$, and $f_{\rm halo}{(\rm all)} = 0.4$, respectively.}} 
        \label{fig:TNG_intrinsic_kinematic_distribution}
    \end{figure}
    
\subsection{Reclassification of fast and slow rotators through the kinematic bimodality}\label{lab_classity_by_GMM}
    
    As established in the Section \ref{label_scaling_krot_spheroid}, \textcolor{black}{$\lambda_{R,\rm{intr}}$} exhibits strong correlations with both $f_{\text{spheroid}}$ and $\kappa_{\text{rot}}$, confirming its utility as a fundamental parameter for characterizing galactic rotation and classifying slow versus fast rotators.

    Figure~\ref{fig:TNG_intrinsic_kinematic_distribution} shows a pronounced bimodality in kinematic properties, with well-separated peaks for fast and slow rotators. Both the $\text{ATLAS}^{\text{3D}}$ and MaNGA surveys exhibit this dichotomy: 
    \begin{itemize}
        \item \text{Fast rotators (FRs) / disk galaxies} are characterized by high intrinsic spin ($\lambda_{{R,\rm intr}}(R_e) \sim 0.7$), thus strong rotational support ($\kappa_{\text{rot}} \sim 0.7$) and low spheroid fractions ($f_{\text{spheroid}} \sim 0.3$).  
        \item \text{Slow rotators (SRs) / elliptical galaxies} have low spin ($\lambda_{{R,\rm intr}}(R_e) \sim 0.2$), weak rotation ($\kappa_{\text{rot}} \sim 0.3$), and dominant spheroid components ($f_{\text{spheroid}} \sim 0.8$). 
    \end{itemize}
    The vertical black dashed lines in Figure~\ref{fig:TNG_intrinsic_kinematic_distribution} mark the new thresholds we suggest for distinguishing slow and fast rotators, giving \textcolor{black}{$\lambda_{{R,\rm intr}}$} $\sim 0.4$, $\kappa_{\text{rot}} \sim 0.5$, and $f_{\text{spheroid}} \sim 0.6$. These thresholds roughly correspond to the minimum probability of the bimodal distributions. The threshold \textcolor{black}{$\lambda_{\rm{R,intr}}$} $\sim 0.4$ is exactly same to that suggested by \citet{Wang_2024}. This bimodal distribution persists across all mass ranges, revealing a significant population of rapidly rotating ETGs. Notably, the $\kappa_{\text{rot}} \sim 0.5$ threshold - commonly used in simulations to select disk galaxies - proves observationally robust for identifying fast rotators and disk galaxies. We maintain the conventional designations of slow and fast rotators but introduce refined kinematic thresholds to better distinguish between these populations. This update ensures consistency with established nomenclature while improving the physical accuracy of the classification scheme. \par 
    
    \textcolor{black}{The distribution of spin parameter ($\lambda_R$) is shown in the second row of Figure~\ref{fig:TNG_intrinsic_kinematic_distribution}. As $\lambda_R$ constitutes the projected counterpart of intrinsic $\lambda_{R,\rm intr}$, \textcolor{black}{it follows that $\lambda_R \sim \lambda_{R,\rm intr} \sin i$ when assuming the galaxy as an isotropic thin disk \textcolor{black}{(see Equation \ref{eq_lambda_re_intr}).}} This systematic projection effect depresses observed $\lambda_R$ values below their intrinsic values. For fast rotators, the peak shifts from an intrinsic value of $\lambda_{R,\rm intr}\sim 0.65$ to a lower observed value near $\lambda_R\sim 0.55$. The threshold decreases to 0.3. It further weakens the bimodality in the distribution and reduces the diagnostic power of $\lambda_R$ alone, making it less accurate than $\lambda_{R,\rm intr}$ for distinguishing between fast and slow rotators. Furthermore, \citet{2021MNRAS.505.3078V} selected ETGs in the SAMI Galaxy Survey using visual morphology, reporting a similarly bimodal distribution of $\lambda_{R_{\rm e}}$ values. As shown in their Figure 8, these bimodality peaks occur near $\lambda_R(R_e) \sim$ 0.5 and $\lambda_R(R_e) \sim $ 0.1 within a comparable stellar mass range to our sample - a result consistent with our findings from both MaNGA and ${\rm ATLAS}^{\rm 3D}$.}\par 

    Figure \ref{fig:TNG_intrinsic_kinematic_distribution} clearly shows a distinct population divergence between FRs classified according to our new threshold (\textcolor{black}{thick black solid profiles in Figure ~\ref{fig:TNG_intrinsic_dynamical_auto_GMM}}) and those identified using the conventional criterion (\textcolor{black}{thin black solid profiles in Figure~\ref{fig:TNG_intrinsic_dynamical_auto_GMM}}). Under the previous classification scheme, only galaxies with minimal disk components and weak rotational support—specifically elliptical systems meeting the conditions $\kappa_\text{rot} < 0.4$, $f_\text{spheroid} > 0.8$, and $f_\text{halo} > 0.6$—were categorized as SRs. The TNG simulations exhibit approximately 10\% more SRs compared to observational samples. Nevertheless, the overall fraction of SRs in TNG remains consistent with those derived from both the $\text{ATLAS}^{\text{3D}}$ and MaNGA surveys. Moreover, the proposed threshold offers a universal criterion for classifying galaxies of all types—including both early-type and late-type galaxies—based on their rotational properties.
    
 \begin{figure*}[t]
        \centering        \includegraphics[width=0.98\textwidth]{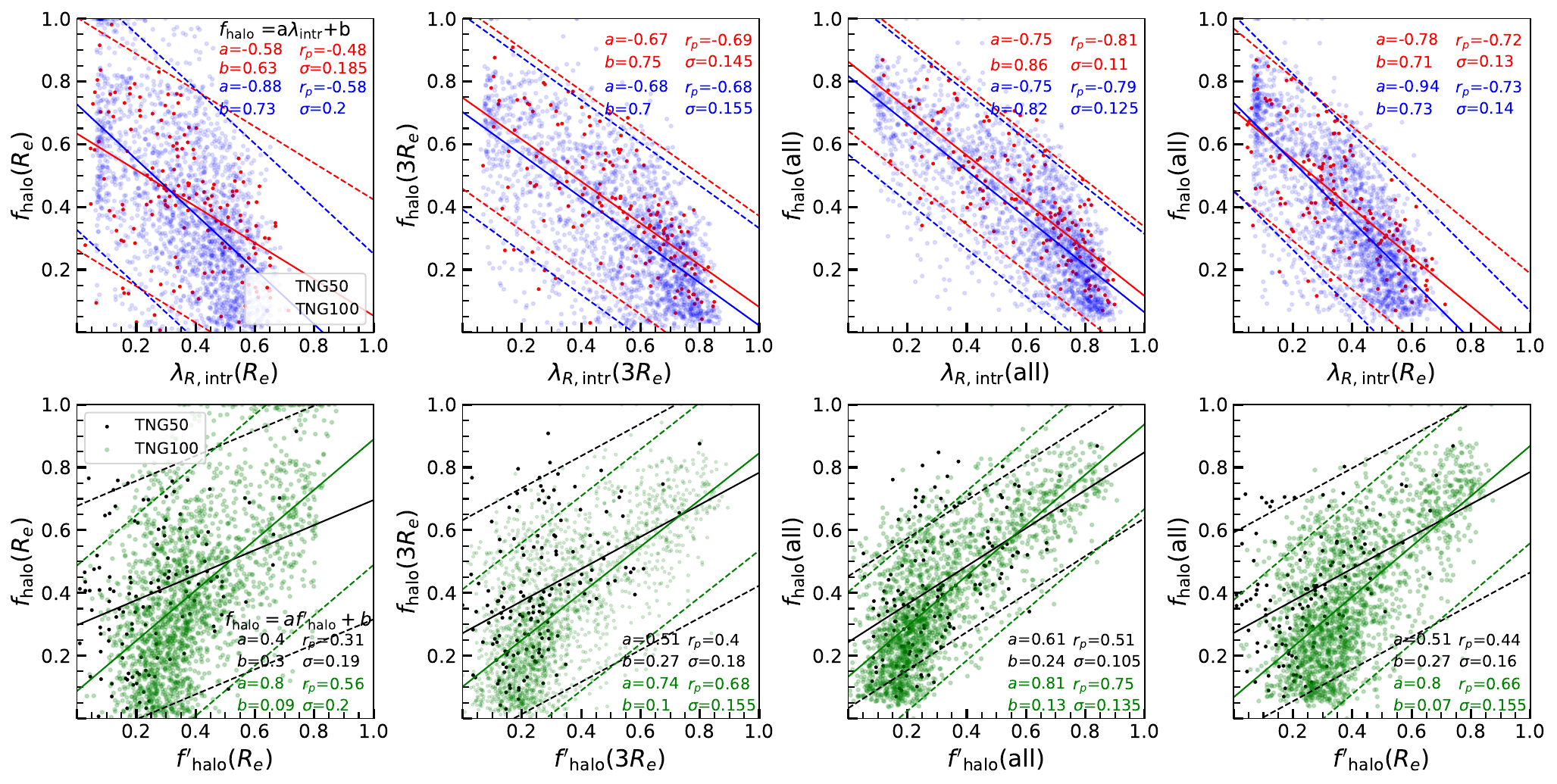}
            \caption{Correlations of \textcolor{black}{$\lambda_{R,\rm{intr}}$} versus $f_{\rm halo}$ and \textcolor{black}{$f'_{\rm{halo}}$} versus $f_{\rm halo}$. The first row presents the relationships between \textcolor{black}{$\lambda_{R,\rm{intr}}$} and $f_{\text{halo}}$ measured within $R_e$ (first column), $3R_e$ (second column), and for all stars (third column). The fourth column shows the relation between \textcolor{black}{$\lambda_{{R,\rm intr}}(R_e)$} and $f_{\text{halo}}(\text{all})$. Solid red and blue lines, along with their corresponding dashed lines, represent the linear fits and $2\sigma$ \textcolor{black}{confidence interval} for the TNG50 and TNG100 datasets, respectively. Fit coefficients, $\sigma$ values, and Pearson correlation coefficients ($r_p$) are provided in the top-right corner of each panel. The second row follows a similar layout but illustrates the relationship between $f'_{\text{halo}}$ and $f_{\text{halo}}$. In this row, solid black and green lines, with accompanying dashed lines, indicate the linear fits and $2\sigma$ \textcolor{black}{confidence interval} for the TNG50 and TNG100 datasets, respectively.} 
        \label{fig:TNG_intrinsic_lambdare_vs_dynamical_3Re_halo_bulge}
    \end{figure*}
   
   \subsection{Kinematic bimodality deficiency in TNG: Systematic under-rotation of simulated ETGs}\label{sec_bimodality}

   While the TNG simulations reproduce a similar overall proportion of SRs to observations—as detailed in Section \ref{lab_classity_by_GMM}—they exhibit a less distinct kinematic bimodality. This discrepancy arises from an overabundance of galaxies with intermediate rotational support, as shown in Figure \ref{fig:TNG_intrinsic_kinematic_distribution}. Both TNG50 and TNG100 show a significant deficit of fast rotators, leading to a systematic shift toward lower intrinsic spin parameter (\textcolor{black}{$\lambda_{R,\rm{intr}}$}) values throughout the simulated population. This deficiency of fast-rotating ETGs in TNG simulations results in systematically different kinematic properties compared to observations: simulated ETGs exhibit weaker rotation (lower $\kappa_{\text{rot}}$), higher spheroid fractions ($f_{\text{spheroid}}$), and higher stellar halo fractions ($f_{\rm halo}$), as shown in Figure~\ref{fig:TNG_intrinsic_kinematic_distribution}. This result manifests two primary shortcomings: (1) a near-total absence of rapidly rotating systems with \textcolor{black}{$\lambda_{{R,\rm intr}}$} > 0.6, and (2) a displacement of the characteristic \textcolor{black}{$\lambda_{R,\rm{intr}}$} peak from the observed value of $\sim 0.7$ to below $0.5$. Correspondingly, the simulations contain an overabundance of slow-rotating ETGs and yield a higher fraction of ETGs with significant spheroidal components ($f_{\text{spheroid}} > 0.5$), many of which represent disk galaxies embedded within massive stellar halos \citep{2025A&A...699A..99H}.
   
    It is worth highlighting that the $\text{ATLAS}^{\text{3D}}$ survey contains approximately $16\%$ fewer massive galaxies compared to MaNGA (27\% vs. 43\%), as evidenced by the sparse population in the top-right corner of Figure~\ref{fig:TNG_intrinsic_kinematic_distribution} and the stellar mass distribution shown in Figure~\ref{fig:mass_size_relation} (top panels). This disparity likely reflects MaNGA's selection bias toward more massive systems, but this issue has no clear effect on the kinematic bimodality, as shown in Figure~\ref{fig:TNG_intrinsic_kinematic_distribution}. TNG100 and TNG50 reproduces the $\text{ATLAS}^{\text{3D}}$ mass distribution reasonably well, though TNG100 slightly overproduces galaxies in the $10^{10.4}-10^{10.8}M_\odot$ range and underrepresents the less massive ETGs. The less pronounced kinematic bimodality is thus unlikely to be caused by the discrepancy of the proportion of ETGs within each mass range relative to the total ETG sample.\par 

    This systematic discrepancy suggests that TNG simulations produce galaxies with higher spheroid and stellar halo mass fractions than those observed. \textcolor{black}{It is still unclear what reasons cause this difference. It} may stem from an insufficient numerical resolution to accurately capture structural properties and their dynamical evolution, particularly in central regions, such as bars, gas inflows, and other nuclear \textcolor{black}{fast-rotating} structures. \textcolor{black}{Furthermore, the treatment approaches of the interstellar medium and feedback subgrid models also have an impact on the kinematics in especially the inner regions of galaxies. Active galactic nucleus and stellar feedback shape the star formation history and even kinematic features via regulating gas cooling and quenching \citep{2018MNRAS.479.4056W,2021MNRAS.500.4004D}.} Although TNG simulations successfully reproduce many galactic properties, accurately matching observed kinematic distributions remains challenging. It is also important to note that observational IFU measurements are typically limited to central regions (within approximately $R_e$), which cannot fully represent the overall rotational characteristics of galaxies and introduce significant uncertainties. Consequently, the apparent bimodality observed in kinematic distributions may also be subject to reliability concerns. \textcolor{black}{Moreover, as mentioned in Section \ref{sample select sec}, TNG50 appears to produce a smaller proportion of ETGs (27\%) compared to observations (33\%), this result also indicates that disk galaxies in TNG simulations exhibit lower quenching probabilities into early-type morphologies compared to observations}. Consequently, in the real universe, systems with significant disk components appear more susceptible to quenching processes that transform them into ETGs—a transition that current simulations may not fully capture.\par

\subsection{A novel method for approximating stellar halo mass fractions using \texorpdfstring{\textcolor{black}{$\lambda_{R,\rm{intr}}$}}{}}

    The ubiquitous presence of massive stellar halos fundamentally challenges conventional galaxy classification and structural decomposition methods \citep{Gadotti2012, 2020ApJ...895..139D, 2025A&A...699A..99H}. These halos primarily form through merger-driven processes—particularly major mergers and strong tidal interactions—which efficiently disrupt the disks of both progenitor and satellite galaxies \citep{2021ApJ...919..135D}. In halo-embedded disk systems, also called Sombrero-like galaxies, where low-mass disks are embedded within massive stellar halos, traditional morphological analyses often misclassify halo mass as part of the disk structure \citep{2020ApJ...895..139D, 2025A&A...699A..99H}, leading to systematic underestimation. While classical bulges have served as historical proxies for merger activity \citep[e.g.,][]{1977egsp.conf..401T, 2001A&A...367..428A}, TNG simulations demonstrate their inadequacy in capturing the full scope of merger histories \citep{2021ApJ...919..135D}, necessitating more robust diagnostics.\par 

    Stellar halos and suppressed rotation serve as sensitive tracers of galactic merger events, as mergers dynamically convert ordered rotation into random motion, \textcolor{black}{ although some disk will survive from mergers with special orbital configuration \citep{2021MNRAS.507.3301Z}}. This process simultaneously builds up stellar halos and erases disk kinematics. The first row of Figure~\ref{fig:TNG_intrinsic_lambdare_vs_dynamical_3Re_halo_bulge} reveals a moderate anti-correlation between \textcolor{black}{$\lambda_{R,\rm{intr}}$} and the stellar halo fraction $f_{\text{halo}}$ derived from \texttt{auto-GMM} analysis, though with a substantial scatter of $\sigma \sim 0.15$. \par

    We therefore introduce a novel methodology that leverages the \textcolor{black}{$\lambda_{R,\rm{intr}}$}–$f_{\text{spheroids}}$ relation established in Section~\ref{label_scaling_krot_spheroid} to estimate stellar halo masses, despite the considerable associated uncertainty. Our approach computes $f'_{\text{halo}} = f_{\text{spheroids}} - f'_{\text{bulge}}$, where bulges are operationally defined as bright central concentrations according to classical morphological criteria \citep{1936rene.book.....H, 2013seg..book..155B}, without distinguishing between classical and pseudo-bulge subtypes. The bulge mass fraction $f'_{\text{bulge}}$ is approximated by integrating the stellar surface density profile over the central 0–1 kpc region in face-on projections, while $f_{\text{spheroids}}$ is approximated using the scaling relation with \textcolor{black}{$\lambda_{R,\rm{intr}}$}. \textcolor{black}{Meanwhile, we examine that $f'_{\text{bulge}}$ deviates further from $f_{\text{bulge}}$ if $f'_{\text{bulge}}$ is approximated by integrating the stellar surface density profile within the central 0–2 kpc or 0–3 kpc regions.}\par 

    We anticipate improved accuracy with future IFS measurements extending to larger radii, as suggested by the second and third columns of Figure~\ref{fig:TNG_intrinsic_lambdare_vs_dynamical_3Re_halo_bulge}. \textcolor{black}{It is noteworthy that currently, tracer populations such as planetary nebulae can, to a certain extent, already facilitate the extension of IFS measurements to larger radii \citep[e.g.,][]{2018A&A...618A..94P}.} $f'_{\rm halo}$ we obtain have a somewhat acceptable agreement with dynamically derived $f_{\text{halo}}$ values, exhibiting a scatter of approximately $\sigma \sim 0.15$. \textcolor{black}{It is worth mentioning that this method is unable to measure stellar halos with $f'_{\rm halo} \lesssim 0.3$, where the stellar halo is too small. This method enables somewhat statistical analysis of merger histories of ETGs with massive stellar halos using IFS measurements} \textcolor{black}{\citep[e.g.,][]{2013ApJ...768L..28H,2016ApJ...821....5D, 2018MNRAS.474.5300D,2019MNRAS.485.2589M,2021ApJ...919..135D,2022A&A...660A..20Z}}, overcoming limitations inherent in purely morphological identification of stellar halos.\par

    \begin{figure}[t]
        \centering
    \includegraphics[width=0.98\columnwidth]{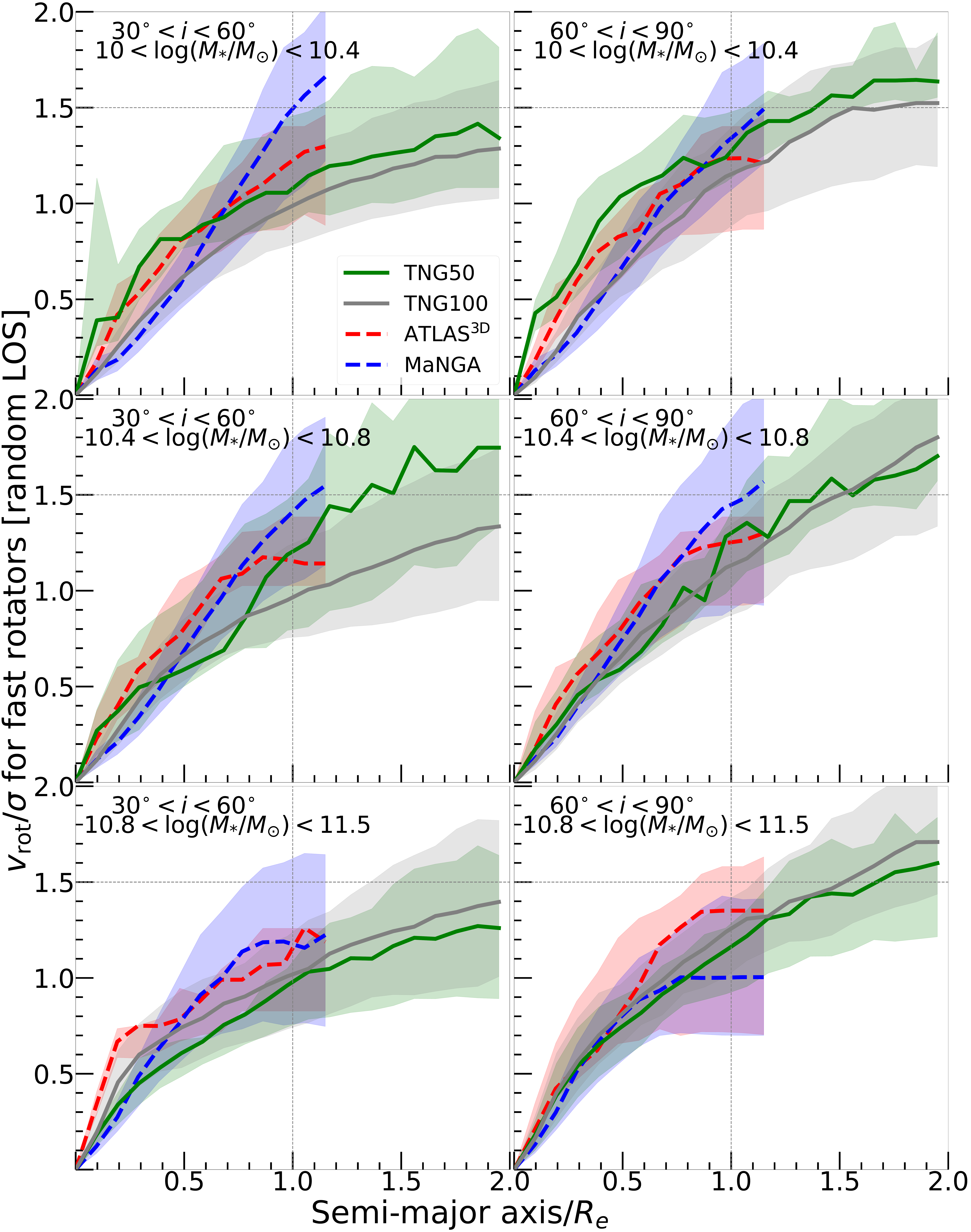}
            \caption{Median $V_{\text{rot}}/\sigma$ profiles for fast rotators selected using our new threshold \textcolor{black}{($\kappa_{\rm rot}(\text{all})>0.5$)}, projected along a random LOS.  \textit{Top row}: it depicts galaxy samples that fall within mass ranges $[10^{10.0},10^{10.4}]M_{\odot}$, from left to right, galaxy samples are presented at different inclinations: $30^{\circ}-60^{\circ}$ and $60^{\circ}-90^{\circ}$, respectively. \textcolor{black}{The green solid lines, gray solid lines, red dashed lines, and blue dashed lines, along with their corresponding shaded regions in the figure, represent the data from TNG50, TNG100, ${\rm ATLAS}^{\rm 3D}$, and MaNGA, respectively.} Both simulation's and observation datasets showcase median profiles, with shaded regions indicating \textcolor{black}{the 16th and 84th} percentiles of the distribution. Vertical and horizontal dashed lines are included as visual aids for comparing simulations and observations. The profiles of simulated ETGs \textcolor{black}{fast rotators} closely resemble those observed. \textcolor{black}{\textit{Middle row} and \textit{Bottom row} are the same as the \textit{Top row}}, \textcolor{black}{but they depict} galaxy samples that fall within mass ranges: $[10^{10.4},10^{10.8}]M_{\odot}$ and $[10^{10.8},10^{11.5}] M_{\odot}$.} 
        \label{fig:rotation_curve_mass_region}
    \end{figure}      
        
\section{Radial distribution of \texorpdfstring{$V_{\rm rot}/\sigma$}{} in the inner regions of selected fast rotators}\label{section rotation}

    Fast-rotating ETGs, characterized by $\lambda_{{R,\rm intr}} > 0.4$ as defined in Section \ref{lab_classity_by_GMM}, serve as ideal tracers of minimally disturbed galactic evolution \textcolor{black}{\citep{2015ApJ...813...23V,2016ARA&A..54..597C,2020MNRAS.495.1958W,2024MNRAS.527..706Z}}. Their kinematic radial profiles, relatively unaffected by major mergers or strong tidal interactions, provide critical benchmarks for studying intrinsic galaxy evolution in both simulations and observations. We thus compare the radial distribution of $V_{\rm rot}/\sigma$ in Figure \ref{fig:rotation_curve_mass_region}. To systematically account for the effects of stellar mass and inclination, we categorize our galaxy sample into three mass bins from left to right. Each mass bin is further divided into two inclination subgroups, i.e., $i=30^{\circ}-60^{\circ}$ and $i=60^{\circ}-90^{\circ}$, from top to bottom. This comprehensive sampling strategy enables us to isolate fundamental kinematic trends while controlling for projection effects and mass-dependent behaviors.\par

    We utilized KINEMETRY to separately fit the galaxy's velocity field $V$ and velocity dispersion field $\sigma$, to obtain the radial distributions of $V_{\text{rot}}$ and $\sigma$ for each galaxy. Figure \ref{fig:rotation_curve_mass_region} shows that the kinematic properties of fast-rotating galaxies in TNG50 (green) match reasonably well with $\text{ATLAS}^{\text{3D}}$ galaxies. TNG100 galaxies exhibit slightly smaller $V_{\rm rot}/\sigma$ values in their central regions, likely due to enhanced numerical heating effects stemming from insufficient resolution, particularly in lower-mass systems. The MaNGA survey similarly shows somewhat depressed $V_{\rm rot}/\sigma$ ratios compared to $\text{ATLAS}^{\text{3D}}$, consistent with expectations given its poorer spatial resolution. 
     \textcolor{black}{Especially, the observed sharp rise in $V_{\rm {rot}}/\sigma$ within the central regions ($R<0.5R_{\rm e}$) of some massive $\rm{ATLAS^{3D}}$ galaxies (bottom-left panel) results from fast-rotating structures. Moreover, it is worth highlighting that $V_{\rm rot}/\sigma$ keeps rising toward $R>R_e$, \textcolor{black}{IFS kinematic measurements} in $R<R_e$ clearly cannot characterize the full dynamical state of fast rotators. The potential \textcolor{black}{existence of massive stellar halos} around disks \citep{Gadotti2012, 2025A&A...699A..99H}, for example \textcolor{black}{the Sombrero} galaxy, may significantly change the overall $\lambda_{R}$ and $V_{\rm rot}/\sigma$ at large radius.}\par

    Our results indicate that the TNG simulations produce a substantially larger population of galaxies with intermediate rotational support, as detailed in Section \ref{lab_classity_by_GMM}. Although TNG50 reproduces the kinematic properties of fast-rotating ETGs—consistent with previous studies \citep{Xu_2019,2025A&A...699A.320Z}—it is important to note that those earlier works did not exclude the potential effect of slow rotators in their samples, a methodological choice that may obscure kinematic distinctions via inducing large scatters. It may diminish the visibility of potential discrepancies. Furthermore, differences in the inclination angles among galaxy samples can also influence the analysis. This overabundance of intermediate-rotators in TNG could explain the shallower \( V_{\text{rot}}/\sigma(R) \) radial profiles reported in \citet{2020A&A...641A..60P}.\par

\section{\textcolor{black}{Discussions: evaluating kinematic analysis in advanced cosmological simulations}}\label{evalute_cosmological_simulation}

    \textcolor{black}{Similar analyses have suggested consistency between TNG simulations and observations. However, when incorporating careful prevalence estimates and inclination corrections, certain discrepancies emerge. The random orientation of galaxies may introduce non-negligible scatter in $\lambda_R-\varepsilon$ diagrams, potentially accentuating apparent simulation-observation agreement—as exemplified by the close but uncorrected alignment shown in Figure \ref{fig:TNG_intrinsic_dynamical_auto_GMM} and \ref{fig:vsigmae}.  \textcolor{black}{The  differences in probability density profiles are discernible (Figure \ref{fig:TNG_intrinsic_kinematic_distribution}, second row).} A typical inclination angle $i$ varies between 30 and 60 degrees, which reduces the spin parameter we measured along the line of sight by a factor of ${\rm sin}(i) \approx 0.5-0.8$. The $\lambda_{R}$ values of fast rotators, generally within the range of 0.3 - 0.7, can thus be significantly underestimated. \textcolor{black}{This geometric effect futher reduces the potential kinematic discrepancies between observations and simulations.} The $\lambda_R-\varepsilon$ diagram thus \textcolor{black}{alone cannot} conclusively establish kinematic consistency between datasets.}\par     
    
    \textcolor{black}{
    Our results highlight the importance of refined analyses when evaluating simulation-observation consistency for not only TNG simulations.  \cite{2018MNRAS.480.4636S}, using the Magneticum Pathfinder simulation, reported general agreement in $\lambda_R(R_e)-\varepsilon$ distributions between simulated and observed ETGs. \cite{2020MNRAS.493.3778S} further extended this approach to $\sim$500 galaxies ($ 2\times 10 ^{10}M_{\odot}<M_{*}<1.7\times 10^{12}M_{\odot}$), noting reasonable matches in two-point gradient relationships. Similar concordance was observed in EAGLE simulations by \cite{2020MNRAS.494.5652W} and \cite{2019MNRAS.484..869V} $(M > 5\times 10^{9}M_{\odot}, z = 0)$, though the latter study noted systematically lower values in Horizon-AGN and Magneticum. Our findings suggest that such comparisons may benefit from explicit treatment of projection effects to ensure conclusive interpretations.}

    \textcolor{black}{Kinematic properties serve as a fossil record of a galaxy's formation history. Consequently, a first-order consistency between simulations and observations may be insufficient, \textcolor{black}{moreover, the non-negligible mixture of slow and fast rotators in the previous classification schemes is also likely to mitigate potential discrepancy.}  In this study, we thus perform detailed statistical comparisons of their projection-corrected distributions across different mass ranges within a unified quantitative framework of both simulations and observations.}

\section{Summary and conclusions}\label{conclusion}
     
    This study presents a comparative analysis of galaxy kinematics between the TNG simulations (TNG50 and TNG100) and observational data from the MaNGA and ATLAS\({}^{\text{3D}}\) surveys. We revisit the classification of early-type galaxies into fast rotators (FRs, disk-dominated systems) and slow rotators (SRs, dispersion-supported ellipticals) using intrinsic dynamical parameters—including the intrinsic spin parameter \textcolor{black}{$\lambda_{R,\rm{intr}}$}, rotational energy fraction \(\kappa_{\text{rot}}\), and spheroid mass fraction \(f_{\text{spheroid}}\). Our results show that widely adopted criteria, such as the empirical \(\lambda_{\rm R}(R_e) = 0.31\sqrt{\varepsilon}\) relation or the Fourier harmonic coefficient \(\overline{k_5}\), are inadequate for robust FR/SR separation due to their weak correlation with fundamental dynamical properties. We instead propose revised thresholds (\textcolor{black}{$\lambda_{R,\rm{intr}}$} $\sim 0.4$, \(\kappa_{\text{rot}} \sim 0.5\), \(f_{\text{spheroid}} \sim 0.6\)) derived from bimodal kinematic distributions, offering a physically motivated universal distinction between rotation- and dispersion-dominated galaxies.\par 

    We also examine inter-parameter correlations among commonly used kinematic quantities and apply scaling relations derived from TNG simulations to estimate \textcolor{black}{$\lambda_{R,\rm{intr}}$}, \(\kappa_{\text{rot}}\), and \(f_{\text{spheroid}}\) in observed MaNGA and ATLAS\({}^{\text{3D}}\) galaxies. Furthermore, we introduce a novel method for inferring stellar halo mass fractions from integral-field spectra (IFS) kinematic measurements, though current estimates remain subject to significant uncertainties.

    Our comparison reveals a notable deficiency in the TNG simulations: they exhibit less distinct kinematic bimodality than observations, producing fewer FRs and showing systematically lower \textcolor{black}{$\lambda_{R,\rm{intr}}$} values. An overabundance of galaxies with intermediate rotational support and elevated spheroid or halo fractions drives this discrepancy. Although the global fractions of SRs and FRs are roughly consistent between simulations ($\sim38\%$ SRs) and observations ($\sim30\%$), the weakened bimodality in TNG reflects a failure to fully capture the dynamical diversity of real early-type galaxies. These findings highlight the need for improved numerical resolution and more sophisticated subgrid physics in future simulations, as well as more extensive observational constraints, to advance our understanding of galaxy formation and evolution.

\section{Data Availability}
    Tables \ref{tab:APPENDIX_lambda_intr_krot_sph_obs} are only available in electronic form at the CDS via anonymous ftp to \url{cdsarc.u-strasbg.fr (130.79.128.5)} or via \url{http://cdsweb.u-strasbg.fr/cgi-bin/qcat?J/A+A/}.
        
\begin{acknowledgements}

    The authors gratefully acknowledge insightful discussions with Luis C. Ho, Lars Hernquist, Bitao. Wang, and Ling. Zhu, \textcolor{black}{and also extend our heartfelt gratitude to the referee for their insightful comments, which have significantly improved the quality of the paper.} This work is supported by the National Natural Science Foundation of China under grant No. 12573010, the National Key R\&D Program of China (No. XY-2025-1459), the Fundamental Research Funds for the Central Universities (No. 20720230015), the Science Fund for Creative Research Groups of the National Natural Science Foundation of China (No. 12221003), and the China Manned Space Program (No. CMS-CSST-2025-A10). The TNG50/TNG100 simulation used in this work, one of the flagship runs of the IllustrisTNG project, has been run on the HazelHen Cray XC40-system at the High Performance Computing Center Stuttgart as part of project GCS-ILLU of the Gauss Centers for Supercomputing (GCS). This work is also strongly supported by the Computing Center in Xi'an, China. The authors also acknowledge the assistance of DeepSeek's AI language model in improving the language and clarity of this manuscript. 
\end{acknowledgements}

\bibliographystyle{aa} 
\bibliography{reference}
\end{document}